\def\CHfour{{\fam0 CH_4}}

\def\COtwo{{\fam0 CO_2}}

\def\CtwoHtwo{{\fam0 C_2H_2}}

\def\HtwoO{{\fam0 H_2O}}
\def\HtwoS{{\fam0 H_2S}}
\def\Htwo{{\fam0 H_2}}

\def\NHthree{{\fam0 NH_3}}

\def\Ntwo{{\fam0 N_2}}

\def\Otwo{{\fam0 O_2}}

\def\cmmthree{{\fam0\, cm^{-3}}}

\def\cmtwo{{\fam0 cm^2}}

\def\scinot#1.{\hbox{$\times 10^{#1}$}}
\def\smone{{\fam0\,s^{-1}}}

\def\ten#1.{\hbox{$10^{#1}$}}

\def\deg{\ifmmode^\circ\else$\null^\circ$\fi}
\def\spose#1{\hbox to 0pt{#1\hss}}
\def\lta{\mathrel{\spose{\lower 3pt\hbox{$\mathchar "218$}}\raise 2.0pt\hbox{$\mathchar"13C$}}}
\def\gta{\mathrel{\spose{\lower 3pt\hbox{$\mathchar "218$}}\raise 2.0pt\hbox{$\mathchar"13E$}}}
\def\lrarrow{\mathrel{\spose{\lower 1pt\hbox{$\rightarrow$}}\raise 3.0pt\hbox{$\leftarrow$}}}

\documentclass{emulateapj}

\slugcomment{submitted to Astrophys. J.}

\shorttitle{Compositional Diversity in the Atmospheres of Hot Neptunes}
\shortauthors{Moses et al.}

\begin{document}


\title{Compositional Diversity in the Atmospheres of Hot Neptunes, with Application to GJ~436\lowercase{b}}

\author{J. I. Moses}
\affil{Space Science Institute, 4750 Walnut Street, Suite 205, Boulder, CO, 80301, USA}
\email{jmoses@spacescience.org}

\author{M. R. Line}
\affil{Division of Geological and Planetary Sciences, California Institute of Technology, Pasadena, CA, 91125, USA}

\author{C. Visscher}
\affil{Southwest Research Institute, Boulder, CO, 80302, USA}

\author{M. R. Richardson}
\affil{Rice University, Houston, TX, 77005-1892, USA}

%
\author{N. Nettelmann} \author{J. J. Fortney}
\affil{Department of Astronomy and Astrophysics, University of California, Santa Cruz, CA, 95064, USA}

%
%
\author{K. B. Stevenson}
\affil{Department of Astronomy and Astrophysics, University of Chicago, Chicago, IL, 60637, USA}

\and

\author{N. Madhusudhan}
\affil{Department of Physics and Department of Astronomy, Yale University, New Haven, CT, 06520-8101, USA}

%
%

\begin{abstract}
Neptune-sized extrasolar planets that orbit relatively close to their host stars --- often 
called ``hot Neptunes'' --- are common within the known population of exoplanets and 
planetary candidates.  Similar to our own Uranus and Neptune, inefficient accretion of 
nebular gas is expected produce hot Neptunes whose masses are dominated by elements heavier 
than hydrogen and helium.   At high atmospheric 
metallicities of 10--10,000$\times$ solar, hot Neptunes will exhibit an interesting continuum 
of atmospheric compositions, ranging from more Neptune-like, H$_2$-dominated atmospheres to more 
Venus-like, CO$_2$-dominated atmospheres.  We explore the predicted equilibrium and 
disequilibrium chemistry of generic hot Neptunes and find that the atmospheric composition 
varies strongly as a function of temperature and bulk atmospheric properties such as metallicity 
and the C/O ratio.  Relatively exotic $\HtwoO$, CO, CO$_2$, and even $\Otwo$-dominated atmospheres 
are possible for hot Neptunes.  We apply our models to the case of GJ 436b, where we find that a 
CO-rich, CH$_4$-poor atmosphere can be a natural consequence of a very high atmospheric metallicity.  
From comparisons of our results with {\it Spitzer\/} eclipse data for GJ 436b, we conclude that 
although the spectral fit from the high-metallicity forward models is not quite as good as the fit 
obtained from pure retrieval methods, the atmospheric composition predicted by these forward models is 
more physically and chemically plausible.   High-metallicity atmospheres (orders of magnitude in 
excess of solar) should therefore be considered as a possibility for GJ 436b and other hot Neptunes.
\end{abstract}



\section{Introduction}

Observations from the {\it Kepler\/} spacecraft reveal that planets with radii between 
that of the Earth and Neptune constitute a dominant fraction of the known transiting exoplanet 
population \citep{borucki11,howard12,batalha13,fressin13}.  Most of these planets and 
planetary candidates orbit relatively close to their host stars and are therefore quite hot 
by Solar-System standards, with equilibrium temperatures $T_{eq}$ $\gtrsim$ 400 K.  These 
so-called ``hot Neptunes'' and hot ``Super Earths'' represent {\em terra incognita\/} for 
planetary researchers.  The atmospheric compositions of these intriguing worlds could be 
widely diverse, ranging from the more familiar hydrogen-dominated ice-giant atmospheres, 
like our own Uranus and Neptune, to relatively hydrogen-poor $\COtwo$-, $\HtwoO$-, or 
N$_2$-dominated atmospheres, or to even more exotic hot metallic, oxygen, and SiO-dominated 
atmospheres, depending on the planet's mass, effective temperature, formation history, 
atmospheric evolution, orbital parameters, and irradiation environment
\citep{elkins08,schaefer09,kite09,rogers10gj1214b,miller-ricci10,miguel11,miller-ricci12,schaefer12,gaidos12,hu12}. 

We investigate the possible atmospheric diversity of hot Neptunes, i.e., close-in transiting 
exoplanets whose radii $R_p$ are typically considered to fall in the $2 R_{\oplus} < R_p < 
6 R_{\oplus}$ range \citep[see][]{borucki11,howard12}, and whose atmospheres contain some H$_2$/He 
component.  Our focus is on how atmospheric properties like temperature, 
metallicity, and bulk elemental ratios can affect the predicted equilibrium and disequilibrium 
composition of the atmospheres of generic hot-Neptune exoplanets, as well as specific hot 
Neptunes such as GJ~436b, for which eclipse observations suggest an unexpected and puzzling 
atmospheric composition \citep{stevenson10,madhu11gj436b}.  

The discovery of GJ 436b by the radial-velocity technique (\citealt{butler04}; see also 
\citealt{maness07}), followed by its identification as a transiting planet \citep{gillon07b}, 
confirmed this intriguing object as the first Neptune-sized exoplanet ever detected.  GJ 436b's 
mass of 1.4$M_{Nep}$ (0.078$M_{Jup}$, 25$M_{\oplus}$), radius of 1.1$R_{Nep}$ (0.37$R_{Jup}$, 
4.1$R_{\oplus}$), and density of 1.2$\rho_{Nep}$ (2.0 g $\cmmthree$), according to \citet{vonbraun12}, 
are all slightly larger than the corresponding values for Neptune.  However, with an orbital 
semimajor axis of only 0.03 AU \citep{torres08,southworth10,vonbraun12}, GJ 436b's dayside 
atmosphere maintains an effective temperature of $\sim$700--900 K 
\citep{deming07,demory07,stevenson10,madhu11gj436b,beaulieu11,knutson11} as a result of 
the strong irradiation from its nearby M-dwarf host star, and thus the planet earns its 
``hot'' designation.
The planet's relatively large radius in relation to its mass suggests that GJ 436b, like 
Neptune, cannot be a purely rocky body or ocean world but must contain a non-negligible 
component of light gases like hydrogen and helium in an outer atmospheric envelope 
\citep[e.g.,][]{fortney07rad,deming07,gillon07b}.  
Interior and evolution models for GJ 436b constrain this H/He component to be $\sim$0.1-22\% by 
mass, depending on other uncertain interior properties and the planet's evolutionary history 
\citep{adams08,baraffe08,figueira09,rogers10frame,nettelmann10,miller11}.

The spectral and photometric behavior of GJ 436b's atmosphere have been studied through 
secondary eclipse observations in the 3.6--24 $\mu$m range 
\citep{deming07,demory07,stevenson10,beaulieu11,knutson11} 
and through transit observations in the $\sim$0.5--8 $\mu$m range 
\citep{gillon07a,gillon07b,deming07,alonso08,bean08,coughlin08,caceres09,pont09,shporer09,ballard10,beaulieu11,gibson11,knutson11}.  
Atmospheric models have been presented by 
\citet{demory07,spiegel10,stevenson10,lewis10,madhu11gj436b,beaulieu11,shabram11} and \citet{line11}.
The consensus from these models is that the large brightness temperatures derived from the {\it Spitzer\/} 
secondary eclipse data are indicative of inefficient heat redistribution from the dayside to the 
nightside of the planet and that the atmospheric metallicity may be greater than solar.  

More contentious are the compositional inferences from the transit and eclipse data for GJ 436b --- 
the transit depths and their implications, in particular 
\citep[cf.][]{pont09,beaulieu11,shabram11,gibson11,knutson11}.  From analyses of transit spectra 
from {\it Hubble\ Space\ Telescope\/} (HST) NICMOS instrument in the 1.1--1.9 $\mu$m range, 
\citet{pont09} and \citet{gibson11} both conclude that the GJ 436b transit spectrum at these 
wavelengths is relatively flat, with no evidence for strong molecular absorption features (including 
those from water); however, the smaller-scale wavelength dependence of the transit depths are notably 
different in the two investigations.  Analyses 
of transit photometric data from the {\it Spitzer\/} IRAC instrument at 3.6, 4.5, and 
8 $\mu$m have led to the conflicting conclusions of a methane-rich atmosphere \citep{beaulieu11} 
and methane-poor atmosphere \citep{knutson11}.  As is discussed extensively in \citet{knutson11}, 
the GJ 436b {\it Spitzer}/IRAC transit depths appear to vary with time, which despite the 
relatively old and apparently quiet nature of the host star, could be due to the occultation of 
star spots or other regions of non-uniform brightness on the star's surface as the planet 
transits across the disk.  The eclipse data are less prone to such problems, although instrument 
systematics are still an issue, and stellar flares can complicate the data analyses 
\citep[e.g.,][]{stevenson12}.  However, some disagreement still exists with respect to the 
inferred planetary flux at 3.6 and 4.5 $\mu$m and the associated error bars at these wavelengths 
in the eclipse data \citep[cf.][]{stevenson10,beaulieu11,stevenson12}, although it is noteworthy 
that there is qualitative agreement in terms of the relative 3.6-to-4.5-$\mu$m flux ratio.

Resolving these discrepancies will be important (and/or obtaining new emission data for GJ 436b) 
because the {\it Spitzer\/} secondary-eclipse data at 3.6, 4.5, 5.8, 8.0, 16, and 24 $\mu$m 
suggest a very unexpected composition for GJ 436b's dayside atmosphere \citep{stevenson10}.  
In particular, the very large flux in the 3.6-$\mu$m IRAC bandpass in combination with the 
negligible flux in the 4.5-$\mu$m bandpass suggest that CO and potentially CO$_2$ are much 
more abundant than CH$_4$ in the atmosphere of GJ 436b (\citealt{stevenson10}; 
\citealt{madhu11gj436b}; but see \citealt{beaulieu11} for a contrary viewpoint), in contrast 
to theoretical models that predict that $\CHfour$ and $\HtwoO$ will be the dominant 
carbon and oxygen constituents.   If CH$_4$ were the dominant carbon-bearing species under GJ 436b 
photospheric conditions, as thermochemical-equilibrium models for solar-like compositions 
suggest \citep[e.g.,][]{lodders02}, then absorption in the 3.6-$\mu$m channel would be stronger 
than is observed, and the brightness temperature of the planet at those 
wavelengths would be much lower.  Invoking a stratospheric temperature inversion in an attempt 
to explain the strong 3.6-$\mu$m emission does not improve the situation because the prominent 
$\nu_4$ band of CH$_4$ would then produce a higher-than-observed flux at 8.0-$\mu$m 
\citep{stevenson10,madhu11gj436b}.  

Through a systematic exploration of parameter space, 
\citet{madhu11gj436b} find that the best fit to all the {\it Spitzer\/} secondary-eclipse 
photometric data is obtained for atmospheres with very high CO mixing ratios, very low CH$_4$ 
mixing ratios, and moderately low $\HtwoO$ mixing ratios, with no thermal inversion.  Both 
\citet{stevenson10} and \citet{madhu11gj436b} suggest that non-equilibrium chemical processes 
could be responsible for this unexpected atmospheric composition, with photochemistry destroying 
the methane in favor of complex hydrocarbons and carbon monoxide, and transport-induced quenching in 
combination with a high-metallicity atmosphere allowing a large quenched CO abundance.  
However, \citet{line11} demonstrate that for assumed atmospheric metallicities up to
50$\times$ solar, photochemistry does not effectively remove CH$_4$ from the GJ 436b's 
photosphere, and transport-induced quenching in combination with photochemistry cannot explain 
the large inferred CO abundance on GJ 436b.  
How then can the puzzling secondary-eclipse observations be explained?

We suggest that other bulk properties of the atmosphere, such as a very high metallicity or 
non-solar elemental compositions, could potentially resolve the current discrepancies 
between models and the secondary-eclipse observations of GJ 436b.  As the atmospheric 
metallicity is increased in a planet with GJ 436b's effective temperature, the overall 
hydrogen mole fraction is decreased, and species like CO and CO$_2$ that do not contain 
hydrogen become progressively favored over hydrogen-containing species like $\HtwoO$ and 
$\CHfour$.  Similarly, as the C/O ratio is decreased, CH$_4$ becomes progressively less 
important in relation to CO and CO$_2$ as a major carbon-bearing constituent 
\citep{madhu12,moses13}.  We note that high-metallicity atmospheres are not unexpected for 
Neptune-mass planets (see section 3.1 below).  In fact, in our own solar system, Neptune's 
atmosphere is observed to have a C/H ratio of 40-120$\times$ solar \citep{baines95,karkoschka11} 
and is estimated to have an O/H ratio greater than 400$\times$ solar \citep{lodders94,luszcz13}. 

We use both chemical-equilibrium models and thermochemical-photochemical kinetics and 
transport models \citep[e.g.,][]{moses11,visscher11,moses13} to investigate ways in which 
CO could be enriched at the expense of CH$_4$ in the atmosphere of GJ 436b.  We also explore 
how bulk properties like atmospheric temperature, metallicity, and the C/O ratio can affect 
the predicted composition of more generic hot Neptunes, leading to potentially widely 
diverse spectral properties for such planets.

\section{Theoretical Models}

\subsection{Chemical Models\label{sectchemmod}}

Two chemical models are used in this study.  The first is a chemical-equilibrium 
model using the NASA CEA code of \citet{gordon94}, and the second is a one-dimensional
(1-D) thermochemical and photochemical kinetics and transport model based on the 
Caltech/JPL KINETICS code of \citet{allen81}.  The kinetics/transport model is described 
more fully in \citet{moses11}, \citet{visscher11}, and \citet{moses13}.  For the equilibrium 
calculations, we consider $\sim$500 gas-phase species and condensates containing the elements 
H, He, C, N, O, Ne, Na, Mg, Al, Si, P, S, Cl, Ar, K, Ca, Ti, Cr, Mn, Fe, and Ni.  For the 
kinetics/transport calculations, we solve the continuity equations for 92 atmospheric species via 
$\sim$1600 forward and reverse chemical-reaction pairs.  Only species containing 
the elements H, He, C, N, and O are considered in the kinetics/transport model due to a lack 
of key rate-coefficient data for species containing the other elements.  Note that the 
included elements are the dominant ones that will not be sequestered within condensates in the 
atmosphere of GJ 436b, and thus the GJ 436b results for the major gas-phase species are not expected to 
change significantly with the inclusion of additional elements.  

The thermodynamic principle of microscopic reversibility, which is expected to be accurate even 
for complex multiple-potential-well chemical reactions \citep{miller09}, is assumed in the 
kinetics/transport model.  Chemical equilibrium is achieved kinetically in the deepest, hottest 
regions of these models, but chemical reactions tend to be slower in the upper, cooler regions.   
When transport time scales drop below kinetic conversion time scales --- at a pressure 
called the ``quench pressure'' or ``quench level'' --- transport begins to dominate over chemical 
kinetics in controlling the species' vertical profiles 
\citep[e.g.,][]{prinn77,lewis84,fegley94,moses10,visscher11}.  When this situation occurs, the 
species can be ``quenched'' at mole fractions that remain constant with altitude above the quench 
level (and thus diverge from chemical-equilibrium predictions), as long as transport times scales 
remain shorter than chemical-kinetics time scales.  Different assumptions about the rate coefficients 
for some key reactions can lead to differences in the predicted abundances of the quenched species 
\citep{visscher11,venot12}; some of the current assumptions in published exoplanet models are 
reviewed by \citet{moses13rev}.  The abundances of the carbon- and nitrogen-bearing species are 
particularly affected when kinetic interconversion between the dominant carbon constituents, CO and 
CH$_4$, and dominant nitrogen constituents, $\Ntwo$ and NH$_3$, ceases to be effective.

Changes to the \citet{moses11} chemical mechanism, other than what is discussed 
in \citet{moses13}, include the addition of O$_3$ and related reactions, which could 
potentially become important as the metallicity is increased.  Ions and ion chemistry 
are not included in the models, and our solutions at the highest altitudes in the model
will be unrealistic due to our neglect of ion chemistry, high-temperature thermospheres, 
and possible hydrodynamic escape \citep[e.g.,][]{garcia07}.  The models contain 198 vertical levels 
separated uniformly in log pressure, with the hydrostatic equilibrium equation being used 
to solve for the background atmospheric parameters along the vertical grid.  The chemical-equilibrium 
abundance profiles from the CEA code are adopted as initial conditions in the kinetics 
and transport models, with zero flux boundary conditions being assumed for all species at 
the top and bottom of the model.  Our assumed solar composition is taken from the 
protosolar abundances listed in Table 10 of \citet{lodders09}.  Multiple Rayleigh scattering 
of incoming stellar radiation by gases is considered in the kinetics/transport models, but we 
assume that aerosols are not present.

We assume that vertical transport occurs through molecular and ``eddy'' diffusion, with the 
eddy diffusion coefficients $K_{zz}$ being free parameters in the model.  In the deep, 
convective portion of the atmosphere, free-convection and mixing-length theories \citep[e.g.,][]{stone76} 
predict relatively large eddy diffusion coefficients and short mixing time scales (e.g., 
$K_{zz}$ $\approx$ 10$^9$ $\cmtwo$ $\smone$ for the convective regions in GJ 436b), 
but $K_{zz}$ values tend to be much smaller in the radiative regions in the upper troposphere 
and lower stratosphere.  Turbulence due to atmospheric tides and upward-propagating gravity 
waves (non-breaking as well as breaking waves) is expected to cause effective $K_{zz}$ values to 
increase roughly with the inverse square root of atmospheric pressure in planetary stratospheres 
\citep[e.g.,][and references therein]{lindzen81} --- a scaling that appears consistent with 
inferred vertical mixing in exoplanet general circulation models (GCMs) (\citealt{parmentier13}; 
cf.~also the $K_{zz}$ profiles inferred from \citealt{showman09}, as shown in \citealt{moses11}).  
In our kinetics and transport models, the $K_{zz}$ profile influences the quench behavior 
of molecules like CO, CH$_4$, and $\NHthree$, and the observations themselves may ultimately 
provide the best means for defining both the $K_{zz}$ values at the quench levels and the pressures 
at which those quench levels occur \citep[e.g.,][]{fegley94,bezard02,viss10co,moses10,visscher11}.  
However, the broadband photometric eclipse observations obtained to date for GJ 436b are not 
sufficient to constrain either the composition or thermal profile accurately enough to derive 
$K_{zz}$ values in such a manner at this time.  We therefore treat $K_{zz}$ as a free 
parameter and will explicitly specify our assumptions for each model presented in section~\ref{sectdisequil}.

\begin{figure}
\includegraphics[angle=-90,scale=0.36]{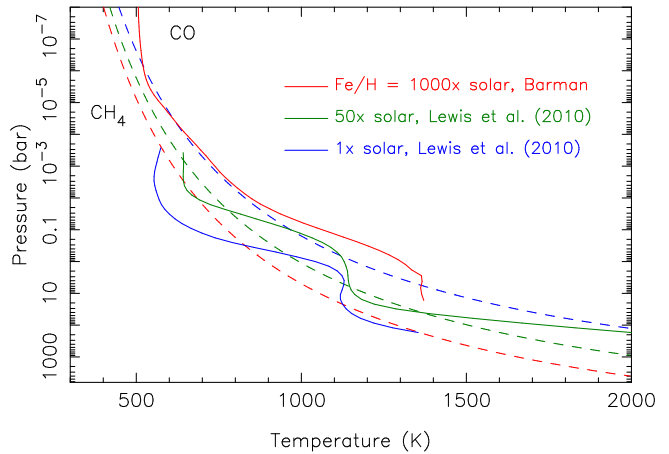}
\caption{Theoretical thermal profiles (solid lines) for GJ 436b assuming various atmospheric 
metallicities.  The profiles for 1$\times$ solar metallicity (blue) and 50$\times$ solar metallicity 
(green) are the atmospheric temperatures averaged over the dayside of GJ 436b at 
secondary-eclipse conditions from the GCMs of \citet{lewis10}; the profile for 1000$\times$ 
solar metallicity (red) is from a 1-D, inefficient-heat-redistribution calculation based on 
\citet{barman05}.  The dashed lines represent the boundaries where CH$_4$ and CO have equal 
abundances in chemical equilibrium for the different metallicity models, with the color coding 
remaining the same as for the thermal profiles.  Methane dominates to the lower left of these 
curves, and CO dominates to the upper right.  A color version of this figure is available in 
the online journal.\label{figtemp}}
\end{figure}

\subsection{Thermal Models\label{sectthermmod}}

Both the chemistry and the predicted spectrum of our hot-Neptune exoplanets will depend 
strongly on the adopted thermal structure.  We do not self-consistently calculate 
temperatures within the kinetics/transport models.  
For our generic hot Neptunes, we explore a wide range of temperature-pressure conditions.  
For GJ 436b, we consider a variety of theoretically-derived thermal profiles, including the 
dayside-average profiles at conditions of secondary eclipse from the 1$\times$ and 50$\times$ 
solar metallicity GCMs of \citet{lewis10}, as shown in Fig.~\ref{figtemp}.   For higher-metallicity 
GJ 436b scenarios, we use the PHOENIX atmospheric model \citep{hauschildt99,allard01} in the 
presence of an external radiation field, as described in \citet{barman01} and \citet{barman05}, 
to compute 1-D temperature-pressure profiles under the assumption of inefficient 
day-night heat redistribution and efficient gravitational settling of condensates (T.~S.~Barman, 
personal communication, 2012).  The resulting
profile for the case of 1000$\times$ solar metallicity is shown in Fig.~\ref{figtemp}.  
For all the non-solar-metallicity thermal models, solar ratios of elements other than hydrogen 
and helium are scaled by a constant factor.  In certain instances, we also adopt thermal 
profiles from \citet{madhu11gj436b} that provide a good fit to the \citet{stevenson10} GJ 436b 
secondary-eclipse data, or we adopt profiles derived from new retrievals such as are described 
in \citet{line12,line13} (see also section~\ref{sectretrieve}).  We typically extend the thermal 
profiles upward in altitude nearly isothermally to a pressure of 10$^{-11}$ bar, where all major 
UV absorbers are optically thin, and downward assuming an adiabat to a lower boundary that 
reaches at least 2400 K, to ensure that the $\Ntwo$--$\NHthree$ quench levels are contained 
within the models (i.e., to encompass the pressure at which kinetic interconversion between 
N$_2$ and NH$_3$ slows down enough to prevent equilibrium from being maintained, see 
\citealt{moses10,moses11}).  The adopted thermal profiles will be clearly described when we 
discuss the different GJ 436b models.

Although the thermal profiles derived from the 3-D GCM and the 1-D PHOENIX models have some notable 
differences for any given metallicity (not shown in Fig.~\ref{figtemp}), both models predict a general 
upward vertical shift in the thermal profile as the metallicity is increased above solar.  An increased 
metallicity results in increased mole fractions of opacity sources like water and other key molecular 
absorbers at lower pressures, which moves the optical-depth unity level (and ``photosphere'' in 
general) upwards to higher altitudes.  However, because the overall atmospheric number 
density decreases with increasing altitude, there is a limit to how high the photosphere can be shifted 
upwards in altitude from this increased metallicity \citep[see also][]{lewis10,miller-ricci10} because
optical thickness cannot be approached at very high altitudes.  As an example, the PHOENIX-based models 
show virtually no differences in the thermal profiles between 1000$\times$ and 10,000$\times$ solar 
metallicities.  The chemical equilibrium CH$_4$-CO equal-abundance boundary also shifts downwards in 
altitude with increasing metallicity (see Fig.~\ref{figtemp}).  

These two metallicity-dependent effects, first noted by \citet{lodders02} and discussed in relation to 
GJ 436b by \citet{lewis10}, motivate our investigation and 
demonstrate why higher-metallicity models are more likely to explain the inferred large CO/CH$_4$ 
ratio needed to reproduce the secondary-eclipse data from GJ 436b \citep{stevenson10,madhu11gj436b}.  For 
instance, Fig.~\ref{figtemp} illustrates that the thermal profile for the 1$\times$ solar metallicity model 
lies completely within the CH$_4$-dominated regime, so that no matter what the rate of vertical mixing 
or at what pressure the CO quenches, the carbon monoxide mole fraction will never exceed the CH$_4$ mole 
fraction for this solar-metallicity thermal profile.  Because the dayside atmosphere is expected to be 
hotter than the terminators, the likelihood of having CO dominate at conditions relevant to the transit 
is even smaller if the atmosphere has a solar-like metallicity.  In contrast, the thermal structure of the 
1000$\times$ solar model shown in Fig.~\ref{figtemp} lies completely within the CO-dominated regime, 
down to at least the 30-bar level.  Since the CO-CH$_4$ quench point (i.e., where transport processes 
dominate over the chemical-kinetic interconversion of CO and CH$_4$) is likely to be within the 
$\sim$0.1-30 bar region on GJ 436b \citep{moses11,visscher11,line11}, the quenching will occur in 
the CO-dominated regime for the 1000$\times$ solar metallicity model, and carbon monoxide will 
dominate over methane.

\begin{figure}
\includegraphics[angle=-90,scale=0.36]{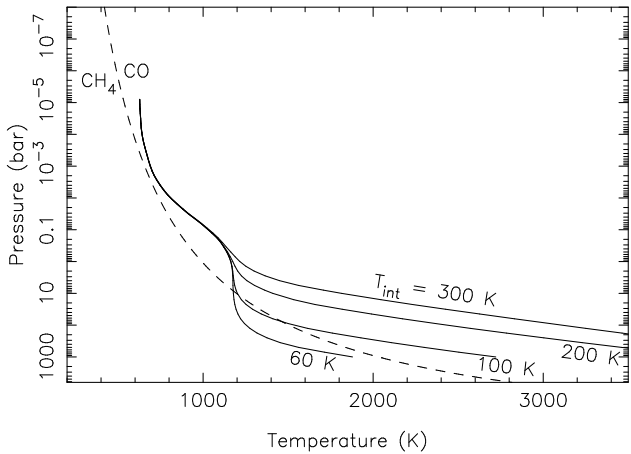}
\caption{Theoretical thermal profiles for GJ 436b assuming different values for the 
intrinsic internal heat flux $F_{int}$ = $\sigma T_{int}^4$, from 1-D, 50$\times$ solar 
metallicity, inefficient heat redistribution calculations that are based on the models of 
\citet{fort06hd149,fortney07rad}.\label{figfortt}}
\end{figure}

In general, the hotter the photosphere of GJ 436b, the more abundant CO is likely to be.  Aside from
the effect of increased metallicities, hotter photospheres can result from the addition of tidal 
heating or a large residual internal heat source.  GJ 436 is a relatively quiet star and slow rotator
\citep{saffe05,demory07,jenkins09,sanzforcada10,knutson10,knutson11}, suggesting 
that the system is relatively old (e.g., 6$^{+4} _{-5}$ Gyr, according to \citealt{torres08}).  For an 
older planet of GJ 436b's mass, the interior would be expected to have cooled significantly such that 
$T_{int}$ $\approx$ 60 K for GJ 436b (comparable to that of Neptune with its $T_{int}$ $\approx$ 50 K), 
defined in terms of an intrinsic internal heat flux of 
$\sigma T_{int}^4$ \citep[e.g.,][]{fortney07rad,marley07,baraffe08,rogers10frame}.  However, 
the orbit of GJ 436b has a significant eccentricity of 0.146 \citep{vonbraun12}, suggesting that the 
atmosphere is being tidally heated.  Since orbital circulation times due to tidal dissipation are 
of order 30 Myr for GJ 436b \citep{deming07}, the eccentricity is likely being continually 
forced by one or more additional planets in the system \citep[e.g.,][]{deming07,demory07,stevenson12}.  
Depending on where the tidal energy is dissipated within the planet, the additional tidal heating could 
increase temperatures at the CO-$\CHfour$ quench point, pushing the thermal profile into the 
CO stability field.  Such a situation is shown in Fig.~\ref{figfortt} for a deep-seated 
intrinsic heat source on a 50$\times$ solar metallicity GJ 436b.  Note, however, that even the 
highest $T_{int}$ profile shown in Fig.~\ref{figfortt} still resides close enough to the CO = $\CHfour$ 
curve in the region from a few bars to a few tenths of a bar that the quenched CH$_4$ abundance is 
likely to be relatively large, even if CO dominates.  A higher metallicity {\it and\/} a large 
internal heat source would make higher CO/CH$_4$ ratios more likely.

\begin{figure}
\includegraphics[angle=-90,scale=0.37]{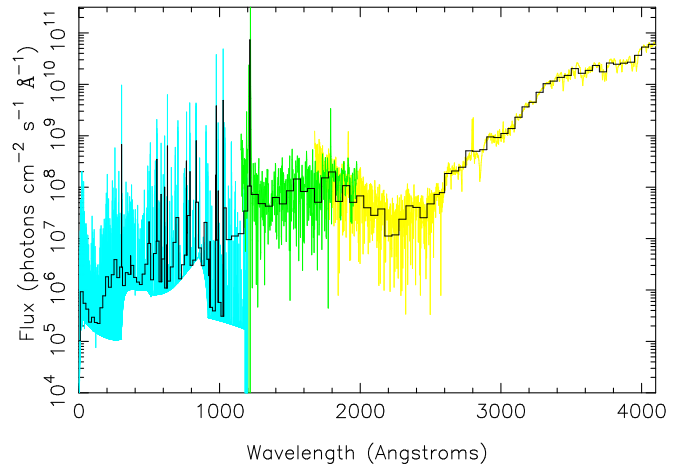}
\caption{Adopted stellar ultraviolet spectrum for GJ 436 (black solid histograms) compared 
to NGSL spectra of GL 15B \citep[yellow; see][]{heap10}, IUE spectra of GL 15B 
(green; {\it Hubble\/} MAST archive), and X-exoplanets theoretical spectra of GJ 436 
\citep[cyan; see][]{sanzforcada11}, all normalized to a distance of 1 AU.  A color version of 
this figure is available in the online journal.\label{fluxstar}}
\end{figure}

\subsection{Stellar Ultraviolet Flux\label{sectstarflux}}

Another input to the kinetics/transport models is the ultraviolet flux from the host star, which has 
been classified in the literature as an early M dwarf of spectral type M2.5{\sc V} to M3.5{\sc V}
\citep{kirkpatrick91,hawley96,maness07,jenkins09}.  Figure~\ref{fluxstar} illustrates our adopted 
flux for GJ 436 (normalized to 1 AU), as derived from a compilation of various observational and 
theoretical sources.  At wavelengths from 0 through the Lyman beta line at 1025.7 \AA, we use the 
GJ 436 synthetic spectrum from the X-exoplanets archive at the Centro de Astrobiolog\'ia (CAB) 
\citep{sanzforcada11}, assuming the units in the downloadable file are photons cm$^{-2}$ s$^{-1}$ 
per wavelength bin, which then reproduces the correct integrated x-ray and EUV luminosity from 
Table 6 of \citet{sanzforcada11}.  At wavelengths between Lyman beta and the Lyman alpha line at 
1215.7 \AA, we use the solar spectrum of \citet{woods02} at solar-cycle minimum, scaled downward 
by a factor of 6.3 to transition smoothly to the longer-wavelength flux.  For the Lyman alpha line itself, 
we use the reconstructed GJ 436 flux from \citet{ehrenreich11}.  For wavelengths between 1215.7 \AA\ 
and 1800 \AA, we use {\it International Ultraviolet Explorer\/} (IUE) data for GL 15B (a M3.5{\sc V} 
star) from the MAST archive (http://archive.stsci.edu), and for wavelengths longer than 1800 \AA, 
we use the spectrum of GL 15B from the Next Generation Spectral Library \citep{heap10}.  For our 
dayside atmospheric models, we scale this normalized 1-AU spectrum to the 0.027 AU orbital distance 
relevant to the GJ 436b secondary eclipse.  The key spectral region as far as the neutral atmospheric 
chemistry is concerned is the wavelength region from Lyman alpha out to $\sim$2400 \AA.  We use a 
fixed solar zenith angle $\theta$ = 48\deg\ to simulate the eclipse conditions in the calculations 
\citep[see][]{moses11}.  A directly measured ultraviolet spectrum for GJ 436 has recently been 
made available by \citet{france13}, and we emphasize that such studies are of great utility to 
theoretical photochemical models for exoplanets.

\subsection{Spectral Models\label{sectradmod}}

To calculate the emergent planetary spectrum for GJ 436b from the assumed thermal structure and 
derived abundance profiles from the chemical models, we use a plane-parallel radiative-transfer 
code as described in \citet{line13}.  Opacity from $\Htwo$, He, $\HtwoO$, CO, CO$_2$, CH$_4$, and 
$\NHthree$ is included in the modeling, although $\NHthree$ itself is not included in the retrievals.  
The source line parameters are described in \cite{line13}.  Local thermodynamic equilibrium is 
assumed throughout, and scattering is ignored.  In a manner similar to \citet{sharp07}, absorption 
cross sections are precomputed using a line-by-line code at high spectral resolution and tabulated 
on a grid with 1 cm$^{-1}$ wavenumber resolution and twenty evenly spaced temperature points between 
500--3000 K and log(pressure) points between 50--10$^{-6}$ bar (see http://www.atm.ox.ac.uk/RFM/).  The 
cross sections from this pre-tabulated grid can then be interpolated for the pressure-temperature-abundance 
conditions relevant to the model atmosphere, which has 90 grid layers equally spaced between 50--10$^{-6}$ 
bar.  As is typical, we plot the model emission spectrum in terms of the flux of the planet divided by 
the flux of the star.  The flux emergent from the entire disk of the planet is calculated using a four-point 
gaussian quadrature.  The stellar flux is derived from a PHOENIX model \citep{hauschildt99,allard01}  
assuming $T_{eff}$ = 3350 K \citep{maness07}, although we should note that the latest determinations 
of the stellar effective temperature from \citet{vonbraun12} are somewhat hotter at $T_{eff}$ = 
3416$ \pm 54$ K \citep[see also][]{bean06,nettelmann10,southworth10}.

\section{Results\label{sectresults}}

Results from our chemical equilibrium and kinetics/transport 
models are presented below.  We first investigate how the predicted equilibrium composition of 
generic hot Neptunes changes as a function of atmospheric temperature, metallicity, and C/O ratio.
We discuss various interesting atmospheric compositional regimes that are not representative of 
planets in our own solar system, and we identify conditions for which CO rather than $\CHfour$ is 
likely to be the dominant carbon component.  We then focus on GJ 436b, determining whether the retrieval 
technique of \citet{line13} can shed any new light on the atmospheric composition of the planet, 
and we calculate disequilibrium chemical abundance profiles for several possible metallicities 
and thermal profiles.  Finally, synthetic spectra from these disequilibrium models are compared with 
{\it Spitzer\/} transit and eclipse data, and observational consequences are discussed.

\begin{figure*}
\begin{center}
\includegraphics[clip=t,scale=0.7]{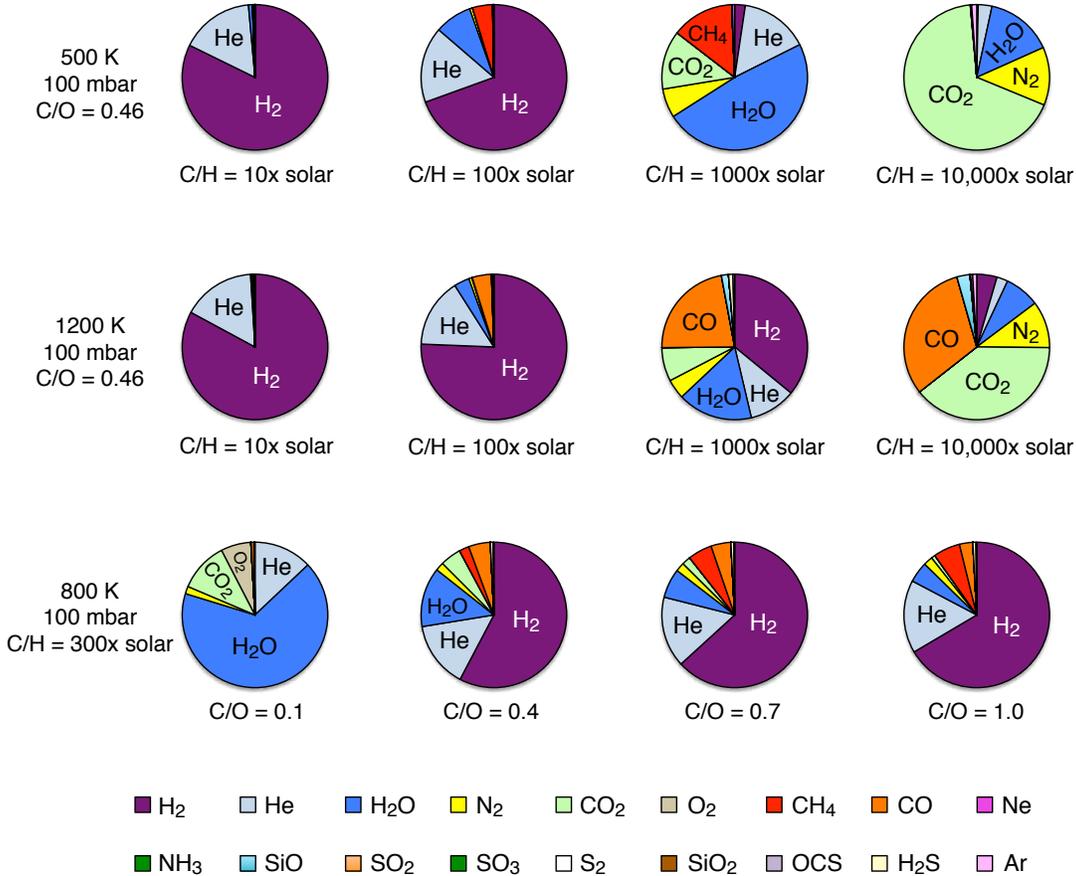}
\caption{Pie charts illustrating equilibrium gas-phase compositions on generic hot Neptunes 
for different assumptions about atmospheric properties.  The top row shows variations as a 
function of metallicity (10-10,000 times solar, as labeled) for an atmosphere with a solar 
C/O ratio, a pressure of 100 mbar, and a temperature of 500 K.  The middle row is similar to 
the top row, except the assumed temperature is 1200 K.  The bottom row shows variations as 
a function of the C/O ratio for an assumed metallicity of 300 times solar (i.e., protosolar 
abundances of all species except H, He, Ne, and O are multiplied by 300, with O being 
defined through the C/O ratio), a pressure of 100 mbar, and a temperature of 800 K.  Note 
the very large variation in composition for different bulk atmospheric properties.
A color version of this figure is available in the online journal.\label{figpie}}
\end{center}
\end{figure*}

\subsection{Chemical Equilibrium for Generic Hot Neptunes\label{sectgeneric}}

With the core-accretion model of giant-planet formation \citep{mizuno78,bodenheimer86,pollack96}, 
a rocky or rock-ice protoplanetary core initially forms and grows from the accretion of solid 
planetesimals within a protoplanetary disk, with gaseous envelopes forming around these emerging 
protoplanets through core outgassing, direct accretion of nebular gas, and ablation of 
incoming solid planetesimals within the gaseous envelope.  Rapid accretion of the surrounding 
largely $\Htwo$ and He nebular gas occurs when the protoplanet reaches a certain critical mass 
($\sim$10 $M_{\oplus}$ in these traditional core-accretion models).  The standard explanation for 
the ``ice giants'' Uranus and Neptune being so much more enriched in heavy elements than Jupiter and 
Saturn is that the accretion rate of solids was slow enough for the proto-Uranus and -Neptune that 
they did not completely reach the runaway gas-accretion-phase before the nebular gas was dispersed from 
the disk \citep[e.g.,][]{lissauer07}.  Elements heavier than hydrogen and helium therefore make up 
80-85\% of Uranus and Neptune by mass \citep{podolak95,hubbard95,fortnet10}.  Depending on the degree of 
mixing between the core and atmosphere, the outermost gaseous envelope could have a heavy-element 
enrichment by mass ($Z_{env}$) different from that of the bulk planet \citep[e.g.,][]{nettelmann13}; 
however, observations of CH$_4$ on the giant planets support the picture of a greater atmospheric 
$Z_{env}$ for the ice giants Uranus and Neptune in comparison with the gas giants Jupiter and Saturn 
\citep[e.g.,][]{moses04,fouchet09,fegley91,gautier95,karkoschka09,karkoschka11}.

A metal-enriched atmospheric envelope might also be true for hot-Neptune exoplanets, despite potentially 
widely different evolutionary and migration histories.  Although the hot Neptunes for which we have good 
constraints on both mass and radius show a large scatter in the mass-versus-radius relation, indicating likely
different compositions, \citet{weiss13} demonstrate that there is a general trend toward increasing 
density with decreasing mass for planetary masses below $\sim150 M_{\oplus}$ \citep[see also][]{miller11}.  
This trend suggests that smaller planets become progressively more enriched in heavy elements, similar to 
the situation in our own solar system.  Planet-formation and population-synthesis models
\citep[e.g.,][]{alibert05,figueira09,mordasini12a,mordasini12b}
also reflect this trend, with \citet{fortney13} predicting a significant increase in $Z_{env}$ of 
Neptune-mass planets in comparison with planets more massive than $\sim 100 M_{\oplus}$.  There 
is also the possibility of {\it in situ\/} formation and capture of gas at small radial distances 
\cite[e.g.,][]{hansen12}, which can lead to a variety of bulk gas fractions depending on planetary 
size and orbital distance.

\begin{figure*}
\begin{center}
\includegraphics[clip=t,scale=0.8]{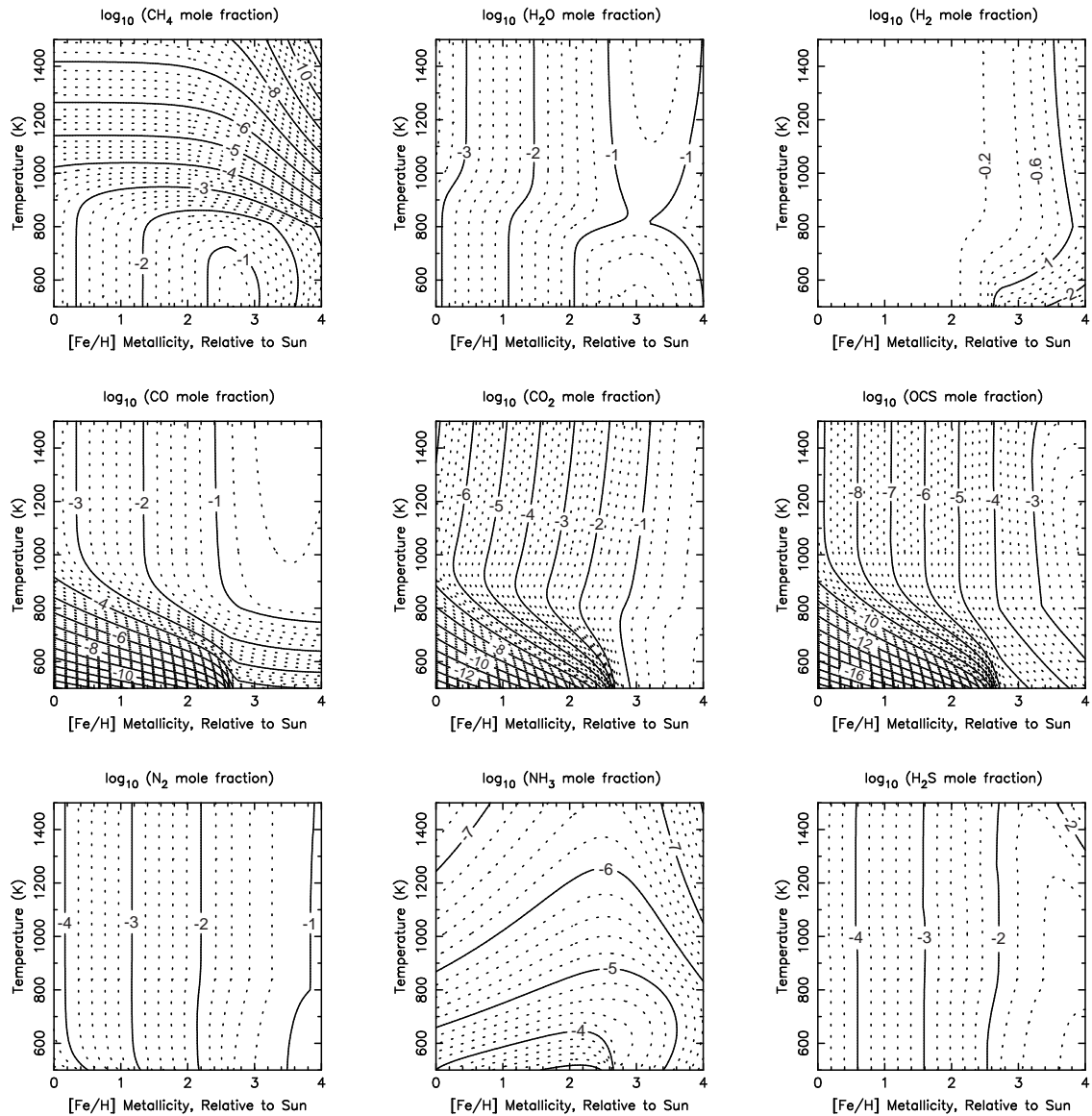}
\caption{Equilibrium mole fractions for different gas-phase species as a function of temperature 
and metallicity for a solar ratio of elements (except H, He, and Ne remain solar at 
all metallicities), at a pressure of 100 mbar.  When we present a metallicity [X/H] in sqaure 
brackets here and in subsequent figures, we use the standard logarithmic definition; e.g., a 
metallicity of [Fe/H] = 3 corresponds to a heavy element enrichment of 1000$\times$ 
solar.\label{figtempvsmetal}}
\end{center}
\end{figure*}

To predict the atmospheric composition of any particular hot Neptune, we would need to know the 
properties of the protoplanetary disk in which the planet formed, the formation location within the 
disk, the planet's migration and impact history, and the details of the subsequent atmospheric 
evolution (e.g., interior outgassing, atmospheric escape, impact delivery/erosion, irradiation 
history, tidal heating, climate evolution, magnetospheric interactions, disequilibrium chemistry, 
etc.).  Given the stochastic nature of some of these evolutionary processes and a lack of information 
about others, the task of predicting any particular atmospheric composition is exceedingly 
difficult (although the attempt can still be made, e.g., \citealt{alibert06}, \citealt{mousis11}, 
\citealt{madhu11carbrich}; moreover, population-synthesis models along the lines of those mentioned 
above can provide valuable insights into atmospheric properties of the ensemble).  Instead of 
pursuing these types of models, we go through the simple exercise of investigating the expected 
equilibrium composition of Neptune-class exoplanets as a function of temperature, bulk metallicity, 
and bulk elemental ratios in the atmosphere.  The increase in metallicity in these calculations 
then becomes a proxy for the evolution of smaller Neptune-mass planets, which are more likely to 
have high $Z_{env}$ due to either inefficient gas accretion or more efficient atmospheric escape.

\begin{figure*}
\begin{center}
\includegraphics[clip=t,scale=0.8]{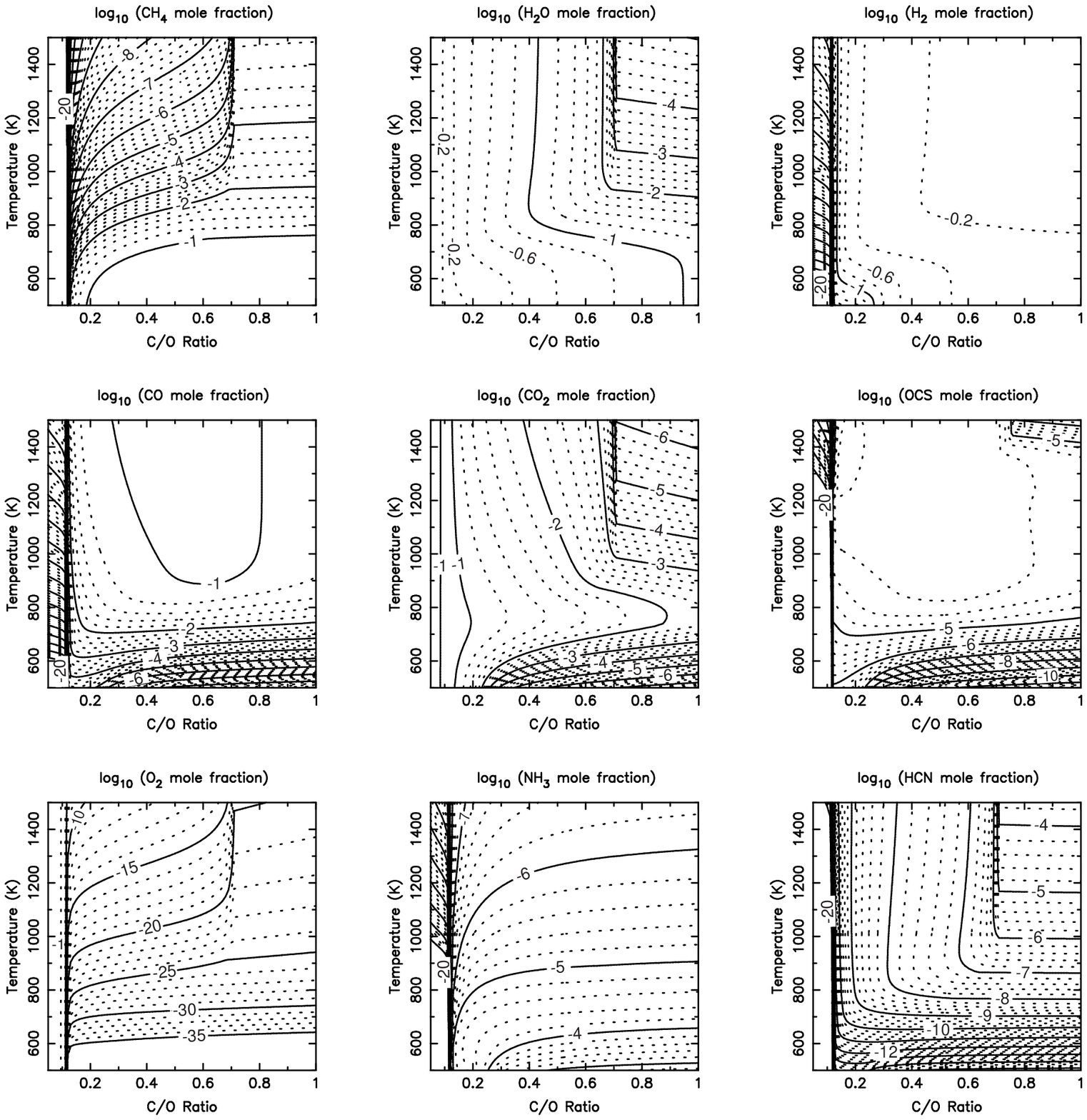}
\caption{Equilibrium mole fractions for different gas-phase species as a function of temperature and 
C/O ratio for a 100-mbar pressure and a metallicity X/H = 300$\times$ solar (where X represents 
all elements except H, He, Ne, and O, with the O abundance being defined through the C/O ratio).\label{figtempvsratio}}
\end{center}
\end{figure*}

Figure \ref{figpie} illustrates how the atmospheric composition is expected to change as a function 
of either the atmospheric metallicity or the atmospheric C/O ratio for a few different temperatures 
at a pressure of 100 mbar.  The 100-mbar pressure was selected here because it represents a typical 
infrared photospheric emission region in transiting-planet atmospheres.  For these calculations, we 
use the protosolar abundances of 
\citet{lodders09} and assume an increase in metallicity occurs uniformly for all species except H, He, 
and Ne; moreover, we define the oxygen abundance through the C/O ratio, so the ``metallicity'' in this 
context refers to the C/H ratio (or X/H ratio, where X is any species except H, He, Ne, or O).  This 
``C/H metallicity'' should not be confused with bulk atmospheric metallicity, as the overall 
heavy-element enrichment will change with the C/O ratio --- at very low C/O ratios, for example, the 
overall atmospheric metallicity, as defined by the abundance of C $+$ O $+$ all heavy elements in 
relation to that of the Sun, can be quite a bit higher than the C/H metallicity due to the exaggerated 
importance of the enhanced O.  Condensates are 
not assumed to rain out in this equilibrium model.  Given such simplifications and given that solar 
ratios of elements may not be preserved as a planetary atmosphere forms and evolves, the relevance 
of these calculations to real planets is questionable, but the exercise does effectively demonstrate 
that the atmospheric composition of hot Neptunes could be highly variable with atmospheric properties.  
Exo-Neptunes with solar-like elemental ratios and moderately low metallicities could have 
hydrogen-dominated atmospheres very reminiscent of our own Neptune, but water and methane will make up 
an increasing fraction of the atmosphere with increasing metallicity, until $\Htwo$ itself becomes 
less important.  At high-enough metallicities, exo-Neptunes can have $\COtwo$-dominated 
atmospheres, qualitatively reminiscent of Venus.  Carbon monoxide becomes an increasingly important 
constituent at higher temperatures and higher metallicities, and even $\Otwo$ can become dominant in 
equilibrium at very high metallicities and very low C/O ratios, further emphasizing that $\Otwo$ by 
itself is not necessarily a good indicator of biological activity on exoplanets 
\citep[e.g., see][]{selsis02,segura07,schaefer12,hu12}.  

\begin{figure*}
\begin{center}
\includegraphics[clip=t,scale=0.5]{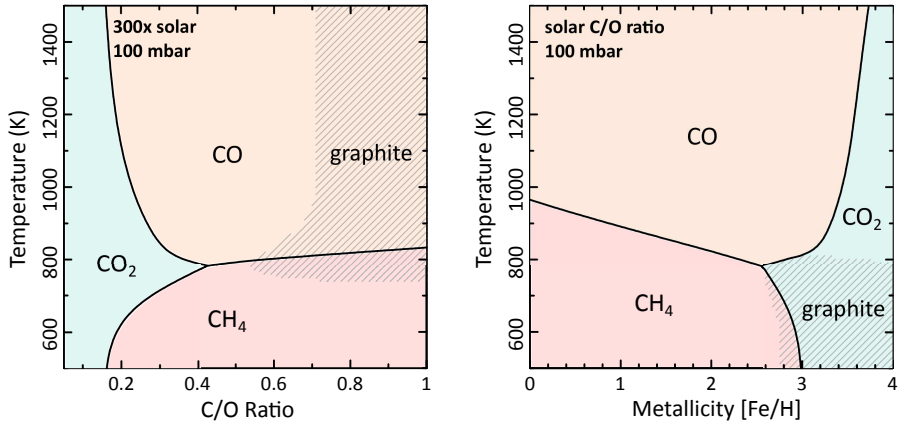}
\caption{Dominant stability regimes in chemical equilibrium at 100 mbar for gas-phase carbon species 
as a function of temperature and bulk C/O ratio for 300$\times$ solar metallicity (left) and as a 
function of temperature and metallicity for a solar C/O ratio (right).  Conditions where graphite is 
stable are shaded.  
A color version of this figure is available in the online journal.\label{figphasebound}}
\end{center}
\end{figure*}

These scenarios are borne out more clearly in Fig.~\ref{figtempvsmetal}, which shows how the 
equilibrium abundances of several atmospheric species vary as a function of temperature and 
metallicity for an assumed pressure of 100 mbar and an assumed C/O ratio that is maintained 
at the protosolar value (C/O = 0.46) with an increase in metallicity (i.e., all elements except 
H, He, and Ne are increased by a constant factor).  Because we do not have a complete set of 
condensates in these equilibrium calculations (and in particular, we are missing several Ca-Ti-Al 
silicates, see \citealt{lodders10}), these gas-phase mixing ratios are not completely accurate, but 
the general trends with temperature and metallicity hold true.  For example, Fig.~\ref{figtempvsmetal} 
demonstrates that $\Htwo$ becomes relatively less abundant at all temperatures as the metallicity 
is increased, whereas the water ($\HtwoO$) mole fraction increases roughly linearly with increasing 
metallicity at most temperatures, until $\HtwoO$ starts to decrease at very high metallicities 
(e.g., $\gta$ 1000$\times$ solar) due to the overall decrease in the bulk H fraction.  Methane 
($\CHfour$) is more stable at low temperatures, where its abundance also increases with increasing 
metallicity until the bulk hydrogen abundance drops and makes less H available to form methane.  
Carbon monoxide (CO) becomes more stable at high temperatures and high metallicities, where it is 
a significant atmospheric component under these conditions.  Carbon dioxide ($\COtwo$) is not very 
abundant at low metallicities, especially at low temperatures, but $\COtwo$ increases significantly 
with increasing metallicity, becoming the dominant constituent at very high metallicities for all 
temperatures.  Carbonyl sulfide (OCS) has a behavior similar to $\COtwo$, although it is never 
as abundant.  Solid graphite is stable in the lower-right corner of these plots in 
Fig.~\ref{figtempvsmetal} (high metallicities, low temperatures) and sequesters a notable 
fraction of the available carbon, thereby lowering the gas-phase C/O ratio in this region of 
temperature-metallicity space.  Ammonia ($\NHthree$) is abundant in equilibrium at the lowest 
temperatures considered, but $\Ntwo$ becomes the most abundant nitrogen component at all 
temperatures as the metallicity is increased, although it never dominates the overall atmospheric 
composition at the conditions considered, assuming no further atmospheric evolution.  The abundance 
of hydrogen cyanide (HCN; not shown in Fig.~\ref{figtempvsmetal}) never rivals that of $\Ntwo$ and 
$\NHthree$, although its abundance increases with increasing temperature at moderately high 
metallicities.  Hydrogen sulfide ($\HtwoS$) is the dominant sulfur constituent under the considered 
conditions, increasingly roughly linearly with metallicity until hydrogen becomes scarce.

Note that atmospheric metallicities of 10, 100, 1000, and 10,000 times solar correspond to 
$Z_{env}$ = 0.13, 0.61, 0.94, and 0.99, respectively, for solar ratios of the full suite of 
elements considered in the models (see section~\ref{sectchemmod}).  This range of $Z_{env}$ 
variation is expected for Neptune-sized planets in the population-synthesis formation and 
evolution models of \citet{fortney13}, with $Z_{env}$ = 0.6-0.9 being the most common for such 
planets.  We might therefore expect the whole range of possible phase space in Fig.~\ref{figtempvsmetal} 
to be represented in the hot-Neptune exoplanet population.  Although Uranus and Neptune are 
good examples of the cool, moderate-metallicity, $\Htwo$-dominated end members, many hot-Neptune 
exoplanets would have atmospheric compositions that are not found in our own solar system, with 
various $\HtwoO$-, CO-, or $\COtwo$-rich possibilities being particularly worth mentioning.
We should also note that metallicities $\gta$ 500--600$\times$ solar are not possible if the 
``metals'' are brought in predominantly via water or very water-rich volatiles (and/or through 
H-rich species such as CH$_4$, NH$_3$, and H$_2$S) --- assuming no further atmospheric evolution 
due to hydrogen escape or other fractionation processes --- because such volatiles would also 
deliver large amounts of hydrogen to the atmosphere \citep{nettelmann11}.

Aside from bulk metallicity, the atmospheric composition is also dependent on the C/O ratio 
\citep{seager05,kuchner05,lodders10,line10,madhu11wasp12b,madhu12,koppa12,moses13}.  
In Fig.~\ref{figtempvsratio}, we show how the equilibrium abundances of several atmospheric 
constituents vary as a function of temperature and C/O ratio for an assumed 100-mbar pressure 
and a moderately-high assumed C/H metallicity of 300$\times$ solar (i.e., where all elements 
other than H, He, Ne, and O are assumed to be 300 times the protosolar abundances of \citealt{lodders09}).  
Recall here that we define the oxygen abundance through the C/O ratio, such that the very low C/O 
ratios at the left edge of these plots also correspond to large overall atmospheric metallicities, in 
addition to O being a dominant element.  For example, our nominal solar C/H ratio is 2.78$\scinot-4.$ 
and O/H ratio is 6.06$\scinot-4.$.  For a uniform enrichment of all elements of 300$\times$ solar, 
the corresponding C/H and O/H ratios would be 8.33$\scinot-2.$ and 1.82$\scinot-1.$, respectively.  
For a C/H ratio of 300$\times$ solar but oxygen defined via the C/O ratio from O/H = (C/H)/(C/O), 
a C/O ratio of 0.08 would correspond to an O/H ratio of 1.04, which is $\sim$1700$\times$ the solar 
O/H ratio.  Under such very low C/O ratio conditions (i.e., C/O $\lta$ 0.08) in this otherwise 300$\times$ 
solar case, hydrogen is no longer the dominant element, and the $\Htwo$ abundance drops precipitously, 
as seen in Fig.~\ref{figtempvsratio}.  Molecular oxygen then becomes the dominant gas at C/O $\lta$ 0.04 
in this scenario --- if such conditions have any relevance to real atmospheres --- but $\HtwoO$ quickly 
takes over with increasing C/O ratio to dominate at 0.05 $\lta$ C/O $\lta$ 0.19 (or at even greater C/O 
ratios for lower temperatures), whereas $\Htwo$ dominates at moderate-to-high C/O ratios (C/O $\gta$ 0.2).  
Carbon monoxide becomes an important constituent at moderate-to-high temperatures and moderate-to-high 
C/O ratios, CH$_4$ and NH$_3$ are important at low temperatures for all but the lowest C/O ratios, 
and $\COtwo$ is important under most conditions except for the lowest temperatures considered at 
high C/O ratios.  Condensed graphite is stable above $\sim$750 K for the highest C/O ratios, and 
HCN increases in significance at high temperatures and high C/O ratios.  Molecular nitrogen (not 
shown in Fig.~\ref{figtempvsratio}) is relatively unaffected by the atmospheric C/O ratio.  

The phase boundaries and stability regimes for the dominant carbon-bearing gases under the conditions 
plotted in Figs.~\ref{figtempvsmetal} \& \ref{figtempvsratio} are further illustrated in 
Fig.~\ref{figphasebound}.  The conditions under which graphite is stable are also shown.  Carbon 
dioxide dominates at very high metallicities and/or very low C/O ratios.  At more moderate metallicities 
and C/O ratios, CH$_4$ dominates at low temperatures and CO at high temperatures.

It is unclear what C/O ratio to expect for hot Neptunes, as that will depend greatly on the planet's 
original formation location (especially in relation to specific condensation fronts in the protoplanetary 
disk), as well as to its evolutionary history \citep[e.g.,][and references therein]{moses13}.  For high 
initial bulk C/O ratios, graphite condensation followed by gravitational settling of the condensates 
(i.e., precipitation) is expected to maintain the gas-phase C/O ratio $\lta 1$ above any graphite clouds
in exoplanet atmospheres \citep{lodders97,seager05,lodders10}; therefore, we do not consider C/O ratios 
$>$ 1 in these models.  However, C/O ratios close to 1 are possible
\citep[e.g.,][]{lodders04,madhu11wasp12b,madhu11carbrich,oberg11,mousis12,moses13} if the planet accreted 
carbon-rich solids inward of the water-ice line in the disk; if the planet accreted CO-rich, $\HtwoO$-poor 
gas from a region between the $\HtwoO$ and CO ice lines, from a region in the innermost disk, or from 
within a heterogeneous disk or water-poor feeding zones; or if the disk were carbon-rich to begin with.  
Similarly, C/O ratios $< 0.1$ can occur if the disk or feeding zones were oxygen-rich or if planetesimals 
composed largely of water ice or clathrate hydrates dominated the delivery of heavy elements to the 
protoplanetary envelope.

The chemical-equilibrium results are also sensitive to pressure, as is shown in \citet{lodders02} 
and \citet{viss06}, for example.  However, given that transport-induced quenching will affect 
the predicted abundances as a function of pressure in real atmospheres, we have simply chosen a 
single representative photospheric pressure for the above figures.  The quench level may occur 
at higher pressures in hot-Neptune atmospheres, thereby affecting the predicted compositions, and 
the infrared photosphere of high-metallicity exoplanets may reside at lower pressures than our 
nominal choice of 100 mbar.  In fact, we emphasize that these simple equilibrium models are largely 
phenomenological and are not designed to represent all the complex processes that have gone into 
shaping the atmospheric composition of actual exoplanet atmospheres.  These equilibrium models do 
serve a useful purpose, though, in illustrating the possible diversity of hot Neptunes and in 
highlighting specific trends such as the increasing dominance of CO$_2$ at very high metallicities, 
the importance of $\HtwoO$ at moderate-to-high metallicities for a variety of other conditions, 
and the change in the relative importance of methane and CO with increasing temperature.  These 
general trends can be useful for considerations of the likely bulk atmospheric properties of 
specific hot Neptunes such as GJ 436b, based on the compositional clues provided by transit and 
eclipse data.

\begin{figure}
\includegraphics[scale=0.38]{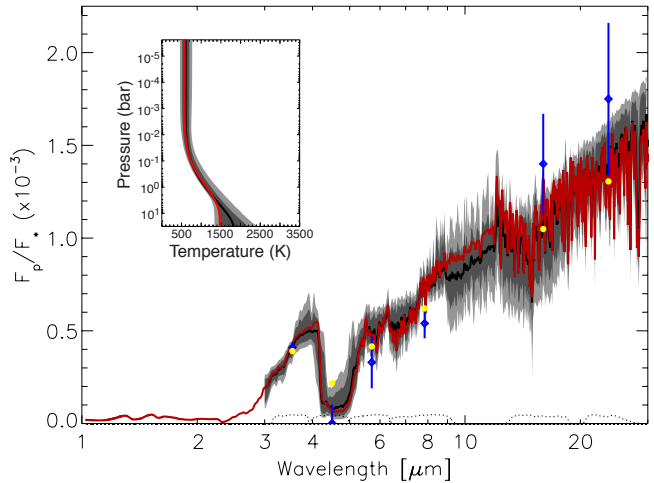}
\caption{Synthetic emission spectra (flux of the planet divided by flux of the star) and thermal 
profile (insert) for GJ 436b derived from the differential-evolution Markov-chain Monte-Carlo 
retrieval method described by \citet{line13}.  The span of solutions that fit within 2-sigma (light 
gray) and 1-sigma (dark gray) are shown in both the temperature-profile and spectral plots, along 
with the median of the ensemble of fits (black curves).  The red curves in both plots represent a 
single best-fit model.  The blue diamonds with error bars are the {\it Spitzer\/} secondary-eclipse 
photometric data from \citet{stevenson10}, and the yellow circles show the best-fit model results 
convolved over the {\it Spitzer\/} bandpasses (with the bandpass sensitivities being plotted as 
dotted curves near the bottom of the plot).
A color version of this figure is available in the online journal.\label{figspecretrieve}}
\end{figure}

\subsection{CHIMERA retrieval methods applied to GJ436b\label{sectretrieve}}

The secondary-eclipse data for GJ 436b (\citealt{stevenson10}; see also \citealt{beaulieu11}, 
\citealt{knutson11}, \citealt{stevenson12}) provide important constraints for compositional 
models.  \citet{madhu11gj436b} use a Markov-chain Monte-Carlo (MCMC) method to help identify 
the range of parameter space allowed for GJ 436b from comparisons of synthetic emission 
spectra with the {\it Spitzer\/} photometric data.  Consistent with the earlier conclusions 
of \citet{stevenson10}, \citet{madhu11gj436b} find that plausible GJ 436b models require a 
low methane abundance (e.g., mole fractions of 10$^{-6}$ to 10$^{-7}$) and a large a CO 
abundance (mole fraction $\ge$ 10$^{-3}$), along with a $\HtwoO$ mole fraction $\le$ 10$^{-4}$ 
and a $\COtwo$ mole fraction in the range $\sim$10$^{-6}$ to 10$^{-4}$ in order to adequately 
reproduce the \citet{stevenson10} eclipse data.  As is noted by \citet{stevenson10} and 
\citet{madhu11gj436b}, such compositions are inconsistent with low-metallicity equilibrium 
models.  \citet{line11} further demonstrate that disequilibrium chemical processes like 
photochemistry and transport-induced quenching do not help resolve this problem.  

\begin{figure*}
\begin{center}
\includegraphics[clip=t,scale=0.78]{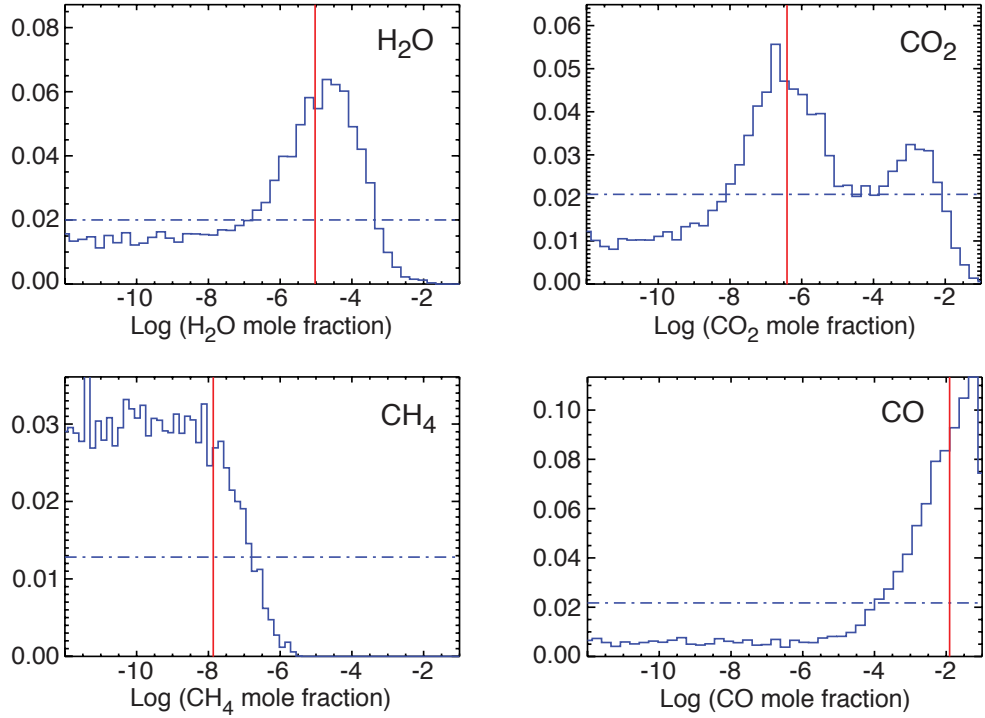}
\caption{Gas mole-fraction histograms of the marginalized posterior probability distribution 
derived from the differential-evolution Markov-chain Monte-Carlo retrieval approach \citep{line13}, 
as applied to the GJ 436b {\it Spitzer\/} eclipse data of \citet{stevenson10}.  The 
horizontal dot-dashed blue curves represent the priors, which are assumed to be flat (uninformative).  
The vertical red lines represent the mole fractions that correspond to the best-fit solution.
A color version of this figure is available in the online journal.\label{fighistogram}}
\end{center}
\end{figure*}

\begin{figure*}
\begin{center}
\includegraphics[clip=t,scale=0.82]{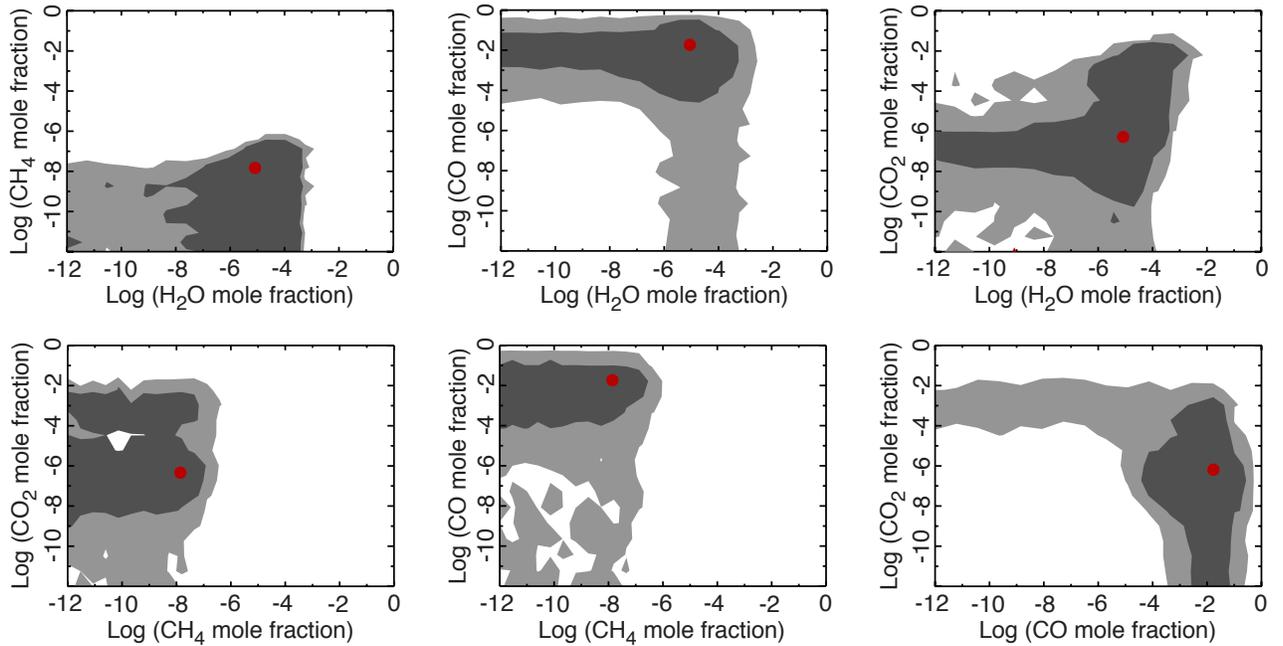}
\caption{Contours of the gas-abundance constraints for GJ 436b derived from the differential-evolution 
MCMC posterior probability distributions, illustrating the correlations between the different gases.  
The dark gray regions represent the 1-sigma confidence interval and light gray regions represent the 
2-sigma confidence interval.  The red dot in each plot is the best-fit (maximum-likelihood) solution 
from the ensemble of 10$^5$ fits.
A color version of this figure is available in the online journal.\label{figcorrelate}}
\end{center}
\end{figure*}

The relatively sparse spectral coverage, low signal-to-noise ratio, and systematic uncertainties
of the {\it Spitzer\/} secondary-eclipse data from GJ 436b and other exoplanets make retrieving 
atmospheric information difficult, as many solutions are statistically valid 
\citep{madhu09,madhu11gj436b,lee12,line12,line13,benneke12,barstow13}.
As a check on the derived best-fit abundances and thermal profile for GJ 436b, we apply 
the CHIMERA retrieval code of \citet{line13} to the \citet{stevenson10} secondary-eclipse data.  
Briefly, CHIMERA employs a suite of Bayesian retrieval algorithms --- optimal estimation, 
bootstrap Monte Carlo, and differential-evolution MCMC --- to determine the allowed range of 
temperatures and gas-phase mixing ratios for GJ 436b.  Here we discuss the results of the 
differential-evolution MCMC approach.  With this technique, the posterior probability distribution 
for each of the parameters that controls the temperature structure (based on the analytic 
parameterizations of \citealt{guillot10}) and assumed constant-with-altitude mole fractions for 
$\HtwoO$, $\CHfour$, CO, and $\COtwo$ is characterized using a genetic algorithm that generates 
$\sim$10$^5$ models \citep[see][for further details]{line13}.  The statistics from the retrieval 
are shown in Figs.~\ref{figspecretrieve}, \ref{fighistogram}, \ref{figcorrelate}.

First, Fig.~\ref{figspecretrieve} shows both the retrieved thermal structure and the corresponding 
spectra for the ensemble of models for GJ 436b generated from the differential-evolution MCMC 
approach.  Rather than plotting the many thousands of spectra and thermal profiles generated from 
this retrieval approach, we instead plot the median of the spectra (or temperatures in the plot 
insert) in black, along with the 1-sigma and 2-sigma spread in the spectra (or temperatures) in 
dark gray and light gray, respectively.  In essence, these spreads reflect the fact that if we 
were to draw a random set of parameters from the posterior probability distributions, there would 
be a 95\% chance that the flux at any one wavelength corresponding to the {\it Spitzer\/} bandpasses 
would fall within the 2-sigma spread, and so on.  Note that although the synthetic spectra qualitatively 
reproduce the 3.6-to-4.5 $\mu$m flux ratio from the {\it Spitzer\/} photometric data, 
the non-detection of the eclipse in the 4.5-$\mu$m bandpass is particularly difficult to reproduce 
\citep[see also][]{stevenson10,madhu11gj436b}.  

Next, Fig.~\ref{fighistogram} shows histograms of the $\HtwoO$, $\CHfour$, CO, and $\COtwo$ mole 
fractions for GJ 436b from the marginalized posterior probability distribution, as derived from 
the differential-evolution MCMC approach.  These distributions show that $\HtwoO$ and CO are 
relatively well constrained from the GJ 436b secondary-eclipse spectra, with the most probable 
$\HtwoO$ mole fraction residing within the range of a few $\scinot-7.$ to a few $\scinot-4.$, and 
the most probable CO mole fraction restricted to $\gta$ 10$^{-4}$, although the solutions for both 
$\HtwoO$ and CO contain an extended, highly unconstrained tail at lower mixing ratios.  The 
methane distribution also shows that the $\CHfour$ mole fraction is restricted to values less 
than 10$^{-6}$.  These results are consistent with the analysis of \citet{madhu11gj436b}.  The 
posterior $\COtwo$ distribution has an interesting double-peaked structure that makes the precise 
value for the $\COtwo$ mole fraction less well constrained, but viable solutions are found for mole 
fractions between 10$^{-8}$ and 10$^{-2}$.  This result for CO$_2$ is also consistent with the 
$\chi ^2$/$N_{obs}$ $\le$ 2 results of \citet{madhu11gj436b}, where $N_{obs}$ is the number of 
available photometric data points.  Our results confirm the picture of a GJ 436b atmosphere that 
possesses a large CO abundance, a very low methane abundance, and a moderately low water abundance.  

Finally, Fig.~\ref{figcorrelate} shows the correlations in the retrieved abundances amongst the 
different gases.  The $\CHfour$ abundance remains low, regardless of the abundance of the other 
species, and thus the $\CHfour$ mole fraction appears uncorrelated with either CO, CO$_2$, or H$_2$O.
An apparent correlation between CO and CO$_2$ results from the very low flux (non-detection) in 
the 4.5-$\mu$m {\it Spitzer\/} channel where both CO and CO$_2$ absorb: if the CO mole fraction 
is large enough, the $\COtwo$ mole fraction is relatively unconstrained and can be low, but if 
the CO mole fraction is small, the $\COtwo$ mole fraction must be large in order to explain the 
very low 4.5-$\mu$m flux.  Note also that solutions with low CO and high CO$_2$ do not fit the 
data as well as models with relatively high CO, perhaps because the high-CO$_2$ solutions end up 
with too much absorption in the 16-$\mu$m {\it Spitzer\/} channel.  Figure~\ref{figcorrelate} 
also illustrates a correlation between $\COtwo$ and $\HtwoO$.  If water is fairly abundant, the 
$\COtwo$ abundance is not well constrained, but if the water abundance is low, the $\COtwo$ 
abundance is more tightly constrained to fall within a mole fraction range of $\sim$10$^{-8}$ to 
10$^{-6}$.  At the highest water abundances, there is also a positive correlation between $\HtwoO$ 
and $\COtwo$ in that very high water abundances require very high $\COtwo$ abundances.  This may 
reflect a strong correlation between $\HtwoO$ (as the gas that has the largest overall influence 
on the spectrum) and temperatures, in that a very large water abundance can only be accommodated 
through high temperatures, at which point more $\COtwo$ is needed to keep the low flux at 4.5 $\mu$m.  
Given that more solutions are found for low-to-moderate water abundances, the low-abundance 
peak for CO$_2$ in Fig.~\ref{fighistogram} dominates over the secondary high-$\COtwo$-abundance 
peak.

These new retrievals do not help resolve the apparent puzzles with respect to atmospheric 
composition on GJ 436b.  Recalling Figs.~\ref{figtempvsmetal} and \ref{figtempvsratio}, a very low 
methane abundance can occur in equilibrium at very high temperatures at low metallicities (i.e., 
$T$ $\gta$ 1200 K in the photosphere, hotter than is likely for GJ 436b) or at more moderate 
temperatures when the C/O ratio is low or when the atmospheric metallicity is very high.  However, 
under those conditions, the $\HtwoO$ abundance is likely to be significantly larger than is indicated 
by all the retrievals \cite[above, and][]{madhu11gj436b}.  Similarly, the preference for high CO 
mole fractions in the retrievals suggests a high metallicity for GJ 436b, which again implies 
unacceptably large water (and potentially CO$_2$) abundances.  In fact, looking at 
Figs.~\ref{figtempvsmetal} and \ref{figtempvsratio}, we need to emphasize that there are no 
equilibrium chemistry solutions --- or even, as we will show below, no disequilibrium chemistry 
solutions --- that fall within the favored temperature-abundance ranges indicated by the retrievals.  
This result illustrates one potential drawback of retrievals from sparse, noisy, systematics-prone data: 
the solutions may reproduce the data well but not make any physical sense.  Going to higher metallicities 
does help in general in that the photosphere is both hotter and more likely to reside in the CO-dominated 
regime (see Fig.~\ref{figtemp}), but a high water abundance is a necessary consequence of high metallicity, 
unless the metallicity is so large ($\gta$ 10,000$\times$ solar) that the atmosphere contains very 
little H.  That solution, which corresponds to $Z_{env}$ = 0.994, (i.e., 0.6\% H/He by mass), is 
unfortunately ruled out by GJ 436b's mass-radius combination, according to most interior models 
\citep[cf.][]{adams08,baraffe08,figueira09,rogers10frame,nettelmann10,miller11}.

\begin{figure*}
\begin{center}
\includegraphics[clip=t,scale=0.8]{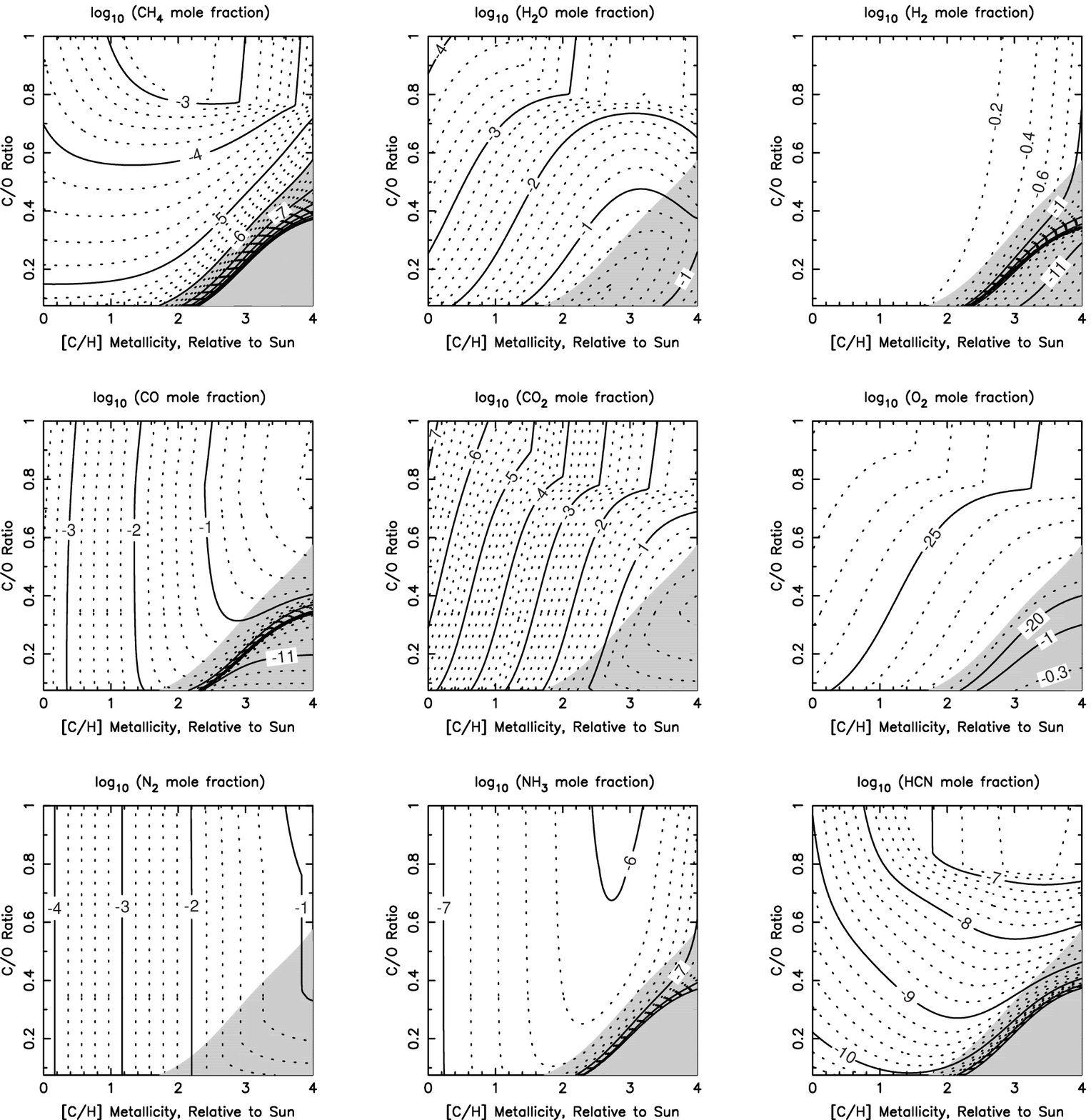}
\caption{Equilibrium mole fractions for different species as a function of C/H metallicity (see text)
and C/O ratio at 900 K and 10 mbar (i.e., a relatively hot, high-altitude photosphere, which may be 
relevant to a high-metallicity GJ 436b).  The shaded region in each plot is where methane has a mole 
fraction below 10$^{-6}$, as is indicated by retrievals based on the \citet{stevenson10} {\it Spitzer\/} 
GJ 436b secondary-eclipse data \citep[our work, and that of][]{madhu11gj436b}.\label{figratiovsmetal}}
\end{center}
\end{figure*}

Figure~\ref{figratiovsmetal} illustrates the possible C/O-ratio vs.~metallicity phase space that 
can accommodate a low methane abundance on GJ 436b for the assumption of a moderately hot, 
high-altitude photosphere (e.g., assuming 900 K at 10 mbar), which may be relevant for a high-metallicity 
GJ 436b atmosphere.  The shaded regions in this plot are consistent with the 
retrieved $\CHfour$ mole fractions of $\lta$ 1 ppm, which can occur at low atmospheric C/O ratios 
and/or high metallicities.  Again, there are no equilibrium soultions within the shaded regions of 
Fig.~\ref{figratiovsmetal} that are entirely consistent with the retrievals, with the main problem 
being too much $\HtwoO$ and too much $\COtwo$.  However, the atmosphere is not likely to be in 
equilibrium \citep{line10,moses13rev}, and we use these constraints as a guide for our disequilibrium 
kinetics and transport models, in an attempt to find forward models with more physically meaningful atmospheric 
properties that can provide a reasonable fit to the GJ 436b {\it Spitzer\/} secondary-eclipse data.

\begin{figure*}
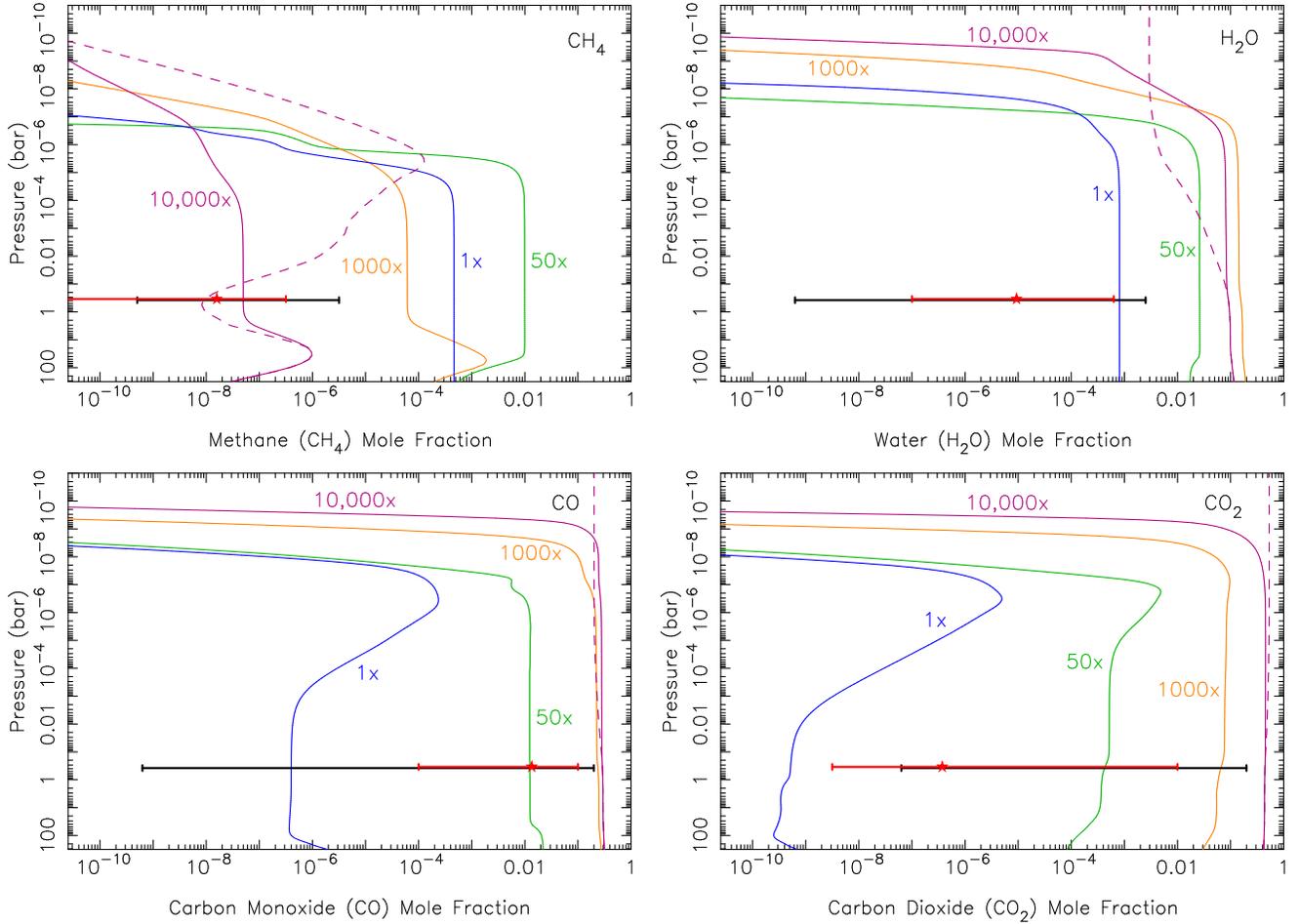

\begin{tabular}{ll}
{\includegraphics[angle=-90,clip=t,scale=0.37]{fig12a_color.ps}}
&
{\includegraphics[angle=-90,clip=t,scale=0.37]{fig12b_color.ps}}
\\
{\includegraphics[angle=-90,clip=t,scale=0.37]{fig12c_color.ps}}
&
{\includegraphics[angle=-90,clip=t,scale=0.37]{fig12d_color.ps}}
\\
\end{tabular}
\caption{Mixing-ratio profiles for $\CHfour$ (top left), $\HtwoO$ (top right), CO (bottom left), 
and CO$_2$ (bottom right) from our kinetics/transport models 
for GJ 436b, for assumed atmospheric metallicities of 1$\times$ solar (blue solid lines), 
50$\times$ solar (green solid lines), 1000$\times$ solar (orange solid lines), and 10,000$\times$ 
solar (purple solid lines), as described more fully in Fig.~\ref{figphotofirst}.  The corresponding 
equilibrium solutions for the 10,000$\times$ solar model are shown as dashed purple lines.  The 
red horizontal bar illustrates the most-probable solutions derived from the differential-evolution 
MCMC approach discussed in section~\ref{sectretrieve}, with the star representing the best-fit solution.  
The black horizontal bar represents the model solutions from \citet{madhu11gj436b} that fit the {\it 
Spitzer\/} data to within $\chi ^2/N_{obs}$ $\le$ 1.  Note that none of the disequilibrium models for 
the different metallicities have mole fractions that fall within the retrieval constraints for all 
four species.  
A color version of this figure is presented in the online journal.\label{figbarmanall}}
\end{figure*}

\subsection{Disequilibrium chemistry modeling of GJ 436b\label{sectdisequil}}

Photochemistry and transport-induced quenching can affect the predicted photospheric abundances 
on extrasolar giant planets, resulting in mixing ratios that are often orders of magnitude 
different from chemical-equilibrium expectations 
\citep{liang03,liang04,zahnle09sulf,line10,line11,moses11,moses13,miller-ricci12,koppa12,venot12,hu12}.  
Most disequilibrium models to date have focused either on giant planets with a near-solar-like 
complement of elements or on hydrogen-poor terrestrial exoplanets or super-Earths.  
We now test various scenarios with moderate and high metallicities (but still containing a non-negligible 
hydrogen mass fraction) for GJ 436b using disequilibrium kinetics/transport models to see how 
photochemistry and transport-induced quenching can alter the predicted transit and 
eclipse spectra.  We start with the first-principles-based temperature profiles shown in 
Fig.~\ref{figtemp} and adopt a solar C/O ratio and a constant eddy diffusion coefficient of 
$K_{zz}$ = 10$^9$ $\cmtwo$ $\smone$ for these initial models.  Fig.~\ref{figbarmanall} shows 
how the model results for $\CHfour$, $\HtwoO$, CO, and CO$_2$ compare with the mixing ratios
retrieved from the {\it Spitzer\/} secondary-eclipse data.  This figure demonstrates that even 
when photochemistry and transport-induced quenching are considered, none of these models can
reproduce the abundances derived from the retrievals.  

\begin{figure*}
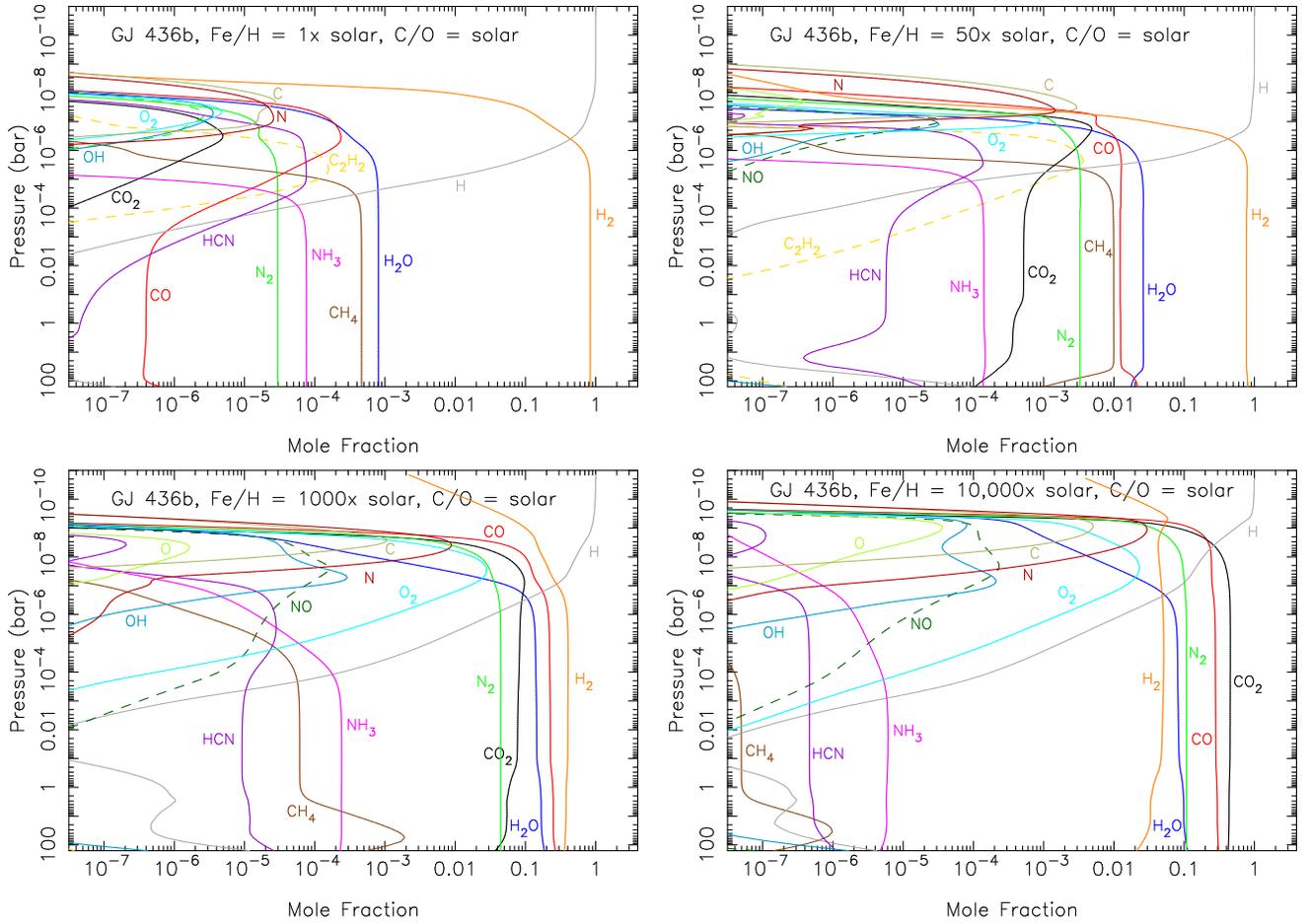

\begin{tabular}{ll}
{\includegraphics[angle=-90,clip=t,scale=0.37]{fig13a_color.ps}}
&
{\includegraphics[angle=-90,clip=t,scale=0.37]{fig13b_color.ps}}
\\
{\includegraphics[angle=-90,clip=t,scale=0.37]{fig13c_color.ps}}
&
{\includegraphics[angle=-90,clip=t,scale=0.37]{fig13d_color.ps}}
\\
\end{tabular}
\caption{Mixing-ratio profiles for several species of interest (as labeled) in our kinetics/transport 
models for GJ 436b, for assumed atmospheric metallicities 
of 1$\times$ solar (top left), 50$\times$ solar (top right), 1000$\times$ solar (bottom left), 
and 10,000$\times$ solar (bottom right).  The thermal profiles adopted in these models are 
described in section~\ref{sectthermmod} (see Fig.~\ref{figtemp}); the profile for the 10,000$\times$ solar model 
(not shown in Fig.~\ref{figtemp}) was derived from the PHOENIX model \citep{hauschildt99,allard01,barman05} 
and is very similar to the 1000$\times$ solar profile.  The C/O ratio is assumed to be the protosolar 
value of \citet{lodders09} with 21\% of the oxygen unavailable due to being bound up in condensates 
at deep atmospheric levels \citep{viss10rock}, and the eddy diffusion coefficient $K_{zz}$ is assumed 
to be constant at 10$^9$ $\cmtwo$ $\smone$.
A color version of this figure is presented in the online journal.\label{figphotofirst}}
\end{figure*}

The mixing-ratio profiles for other potentially interesting species in these disequilibrium models 
are shown in Fig.~\ref{figphotofirst}.  The species profiles 
for the 1$\times$ solar model are qualitatively similar to those of \citet{line11}: behind $\Htwo$ 
and He, the next most abundant gases are $\HtwoO$, $\CHfour$, and $\NHthree$.  Water survives at 
a near-equilibrium abundance until very high altitudes ($\sim$1 microbar), at which point 
photolysis and other destruction mechanisms start to irreversibly convert the $\HtwoO$ to CO and 
other oxygen-bearing species.  Although photolysis and other chemical mechanisms operate to destroy 
$\HtwoO$ at lower altitudes, water is efficiently recycled in the background $\Htwo$ 
atmosphere, and $\HtwoO$ remains the dominant infrared opacity source in these 1$\times$ solar 
models.  Methane is even less stable than water at high altitudes, due primarily to the large H 
abundance released by $\HtwoO$ photolysis and subsequent catalytic destruction of $\Htwo$ 
\citep[e.g.,][]{liang03,moses11}.  The carbon that was in high-altitude $\CHfour$ gets 
photochemically converted to CO, $\CtwoHtwo$, HCN, and atomic carbon, primarily (see \citealt{moses11} 
for details).  Ammonia is photolyzed by longer-wavelength UV radiation that can penetrate a bit
deeper in the atmosphere, and the nitrogen liberated by the $\NHthree$ photodestruction largely 
ends up as HCN and atomic N.  There are some quantitative differences between our 1$\times$ solar 
model and that of \citet{line11} due to different adopted reaction rate coefficients, $K_{zz}$ 
assumptions, stellar ultraviolet flux, and atmospheric temperatures, but the results from both 
\citet{line11} and our own models indicate that $\CHfour$ survives photochemical destruction 
throughout the bulk of the $\sim$0.0001--1 bar photosphere on the 1$\times$ solar GJ 436b.  
As a result, the 1$\times$ solar model has more methane than is indicated by the retrieval 
analyses of the {\it Spitzer\/} secondary-eclipse data (see Fig.~\ref{figbarmanall}, section~\ref{sectretrieve},
and \citealt{madhu11gj436b}).

\begin{figure}
\includegraphics[angle=-90,scale=0.36]{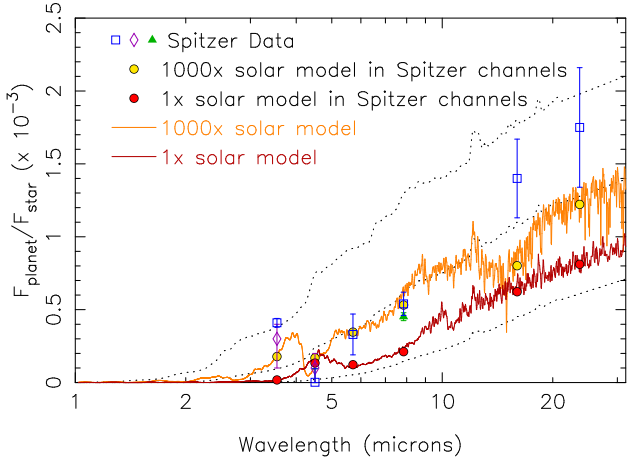}
\caption{Synthetic emission spectra for GJ 436b for our disequilibrium models that assume a 
1$\times$ solar metallicity atmosphere (dark red) and a 1000$\times$ solar metallicity atmosphere 
(orange), in comparison with observations (data points with error bars).  The thermal profiles 
assumed in the models are shown in Fig.~\ref{figtemp}, and the abundance profiles are shown in 
Fig.~\ref{figphotofirst}.  The blue squares represent the {\it Spitzer\/} secondary-eclipse 
data analyzed by \citet{stevenson10}, with an updated upper limit at 4.5-$\mu$m from \citet{stevenson12},  
the purple diamonds represent the \citet{beaulieu11} analysis of the same 3.6 and 4.5-$\mu$m 
data, and the green triangle represents the analysis of 11 secondary eclipses of GJ 436b in the 
8-$\mu$m channel by \citet{knutson11}.  The red and gold circles represent, respectively, the fluxes from 
the 1$\times$ and 1000$\times$ solar models, averaged over the {\it Spitzer\/} bandpasses.  
The black dotted lines represent the planetary emission (smoothed) assuming GJ 436b radiates as a 
blackbody at a temperature of 500 K (lower curve), 800 K (middle curve), or 1100 K (upper curve).  The 
apparent emission ``spikes'' represent stellar absorption, as everything here is plotted in terms of 
the flux of the planet divided by the flux of the star.  A PHOENIX stellar model 
\citep[e.g.,][]{hauschildt99} with $T_{eff}$ = 3350 K was assumed for the host star for all these 
calculations.  A color version of this figure is available in the online journal.\label{figspec1x}}
\end{figure}

The resulting spectrum for the 1$\times$ solar model does not compare well with the {\it Spitzer\/} 
secondary-eclipse data (Fig.~\ref{figspec1x}).  One obvious problem is a reversed 3.6-to-4.5-$\mu$m 
flux ratio in comparison with observations.  The CO-$\CHfour$ quench point in the 1$\times$ solar 
model occurs within the $\CHfour$-dominated regime, so that the quenched CO abundance is relatively 
small, and the methane abundance is large.  Since the equilibrium abundance of CO is everywhere lower 
than that of $\CHfour$ with the 1$\times$ solar thermal profile (see Fig.~\ref{figtemp}), there is 
no eddy $K_{zz}$ value we could adopt that would change this conclusion.  Some CO is produced at high 
altitudes from $\CHfour$ and $\HtwoO$ photochemistry, but the CO column abundance is never large 
enough to influence the relative 3.6-to-4.5-$\mu$m flux ratio.  The large methane abundance therefore 
results in a lower flux (more absorption) at 3.6 $\mu$m in comparison with 4.5 $\mu$m.  Another obvious 
problem is the overall lower brightness temperature at most wavelengths in comparison with the data.  
The thermal profile from \citet{lewis10} adopted in the 1$\times$ solar model is not significantly 
colder than the profiles derived from the thermal retrievals shown in \citet{madhu11gj436b} or in 
section~\ref{sectretrieve}, but the 1$\times$ solar model does have a water abundance at the upper end 
(or greater than) the retrievals indicate (see Fig.~\ref{figbarmanall}).  Water accounts for most of 
the absorption in the 1$\times$ solar model, and the relatively large water abundance leads to more 
absorption at most wavelengths than can be accommodated by the observations -- hence the requirement 
of a low water abundance from the retrievals in the first place.  Since photochemistry does not 
permanently destroy the photospheric water and methane in our 1$\times$ solar models, such models 
cannot explain the {\it Spitzer\/} secondary-eclipse data.

The quench point for the 50$\times$ solar model is closer to the equilibrium CO-CH$_4$ equal-abundance 
boundary due to an overall hotter atmosphere (see Fig.~\ref{figtemp}), and both CO and $\CHfour$ end 
up being major atmospheric constituents (see Fig.~\ref{figphotofirst}).  Carbon dioxide and molecular 
nitrogen also become more important constituents at the higher 50$\times$ solar metallicity, while 
ammonia becomes relatively less important due to the $\Ntwo$-$\NHthree$ quench point falling within 
the $\Ntwo$-dominated regime.  Because $\NHthree$ is more photochemically active than $\Ntwo$, the 
photochemical production of HCN and complex nitriles does not increase significantly at high altitudes 
in the higher-metallicity model because the overall $\NHthree$ mixing ratio has not changed much 
between 1$\times$ and 50$\times$ solar metallicities.  However, the column abundance of HCN actually 
increases with the increased metallicity here because of thermochemical kinetics and the quenching of 
$\CHfour$ and $\NHthree$ at higher-than-equilibrium abundances (i.e., HCN in the photosphere maintains 
a pseudo-equilibrium with the quenched $\NHthree$ and $\CHfour$; see \citealt{moses11,moses13}).  
Carbon dioxide is produced effectively at high altitudes from the photochemistry of CO and $\HtwoO$, 
but equilibrium through the net reaction CO + $\HtwoO$ $\leftrightarrows$ $\COtwo$ + $\Htwo$ is maintained 
kinetically through much of the photosphere.  Although the CO mole fraction in the 50$\times$ solar model 
now falls within the range required by the retrievals discussion in section~\ref{sectretrieve}, both 
$\HtwoO$ and $\CHfour$ are much more abundant than is allowed by the retrievals, and the 50$\times$ 
solar model does not fit the {\it Spitzer\/} secondary-eclipse data well.

At metallicities of 1000$\times$ solar, hydrogen and helium now make up only $\sim$6\% of the 
atmosphere by mass.  The methane abundance has dropped significantly in the 1000$\times$ solar 
metallicity model because the CH$_4$-CO quench point lies within the CO-dominated regime.  The 
resulting CO/$\CHfour$ ratio is finally in the right direction to explain the observed relative 
3.6-to-4.5-$\mu$m flux ratio from the {\it Spitzer\/} eclipse data.  Carbon monoxide begins to 
rival $\Htwo$ as the dominant constituent, and $\HtwoO$, $\COtwo$, and $\Ntwo$ are all very 
abundant.  The atmosphere as a whole is more oxidized, and species like $\Otwo$ and NO that are 
produced from photochemistry at high altitudes are able to survive more readily than they do in 
a more reducing atmosphere.  The fact that elements other than O, C, N, and H have been neglected 
from these disequilibrium models (particularly Cl and S) will have an impact on the resulting 
abundances in these higher-metallicity models due to omitted catalytic cycles \citep[e.g.,][]{yung99}, 
and atmospheric escape and other evolutionary processes will likely alter this simple picture 
of steadily increasing metallicity, but the general supplantation of $\Htwo$ and the increasing 
dominance of $\COtwo$ with increasing metallicity is a robust conclusion.  At atmospheric 
metallicities of 10,000$\times$ solar, for instance, hydrogen and helium make up only $\sim$0.6\% 
of the atmosphere by mass, and Fig.~\ref{figphotofirst} shows that $\COtwo$ has solidly replaced 
$\Htwo$ as the dominant constituent, and even CO, $\Ntwo$, and $\HtwoO$ are more abundant than $\Htwo$.

Figure \ref{figspec1x} shows that the 1000$\times$ solar metallicity model fares better in 
reproducing the {\it Spitzer\/} secondary-eclipse data than the 1$\times$ solar metallicity model 
did.  The flux in the 3.6 $\mu$m channel is predicted to be greater than that in the 4.5 $\mu$m model 
now with the 1000$\times$ solar model, due to the $\CHfour$ abundance being much less than the CO 
abundance.  The 1000$\times$ solar model still provides too much absorption at 3.6 $\mu$m to be 
consistent with the eclipse depth as determined by \citet{stevenson10}, indicating too much methane 
in the model; however, we should note that \citet{beaulieu11} derive both a lower 3.6-$\mu$m 
flux and a larger uncertainty in the eclipse depth at this wavelength (and at 4.5 $\mu$m), and the 
1000$\times$ solar model more readily satisfies those constraints.  More intriguing is the excellent 
fit in the 5.8 and 8.0 $\mu$m channels where water in this model is providing the dominant 
opacity source.  This good fit with a water abundance that exceeds a 10\% mixing ratio --- in contrast 
to the much lower abundances derived from the retrievals --- reflects the influence of the 
higher-temperature photosphere that results from the increased atmospheric opacity at high metallicities.  
The water abundance and photospheric temperatures are closely linked, and a hotter photosphere 
requires more water to remain consistent with the 5.8 and 8.0 $\mu$m eclipse depths, whereas, 
conversely, a colder photosphere requires less water.  Still, this 1000$\times$ solar model 
provides a relatively poor fit to the eclipse data of \citet{stevenson10,stevenson12}, providing 
a $\chi^2 /N_{obs}$ $\sim$ 11.5, with the main problem being a 3.6-$\mu$m flux that is much lower 
than observed, but also a 4.5-$\mu$m flux that is in excess of the observational upper limit and a 
16-$\mu$m flux that is much lower than observed (due to CO$_2$ absorption).

\begin{figure}
\includegraphics[angle=-90,scale=0.36]{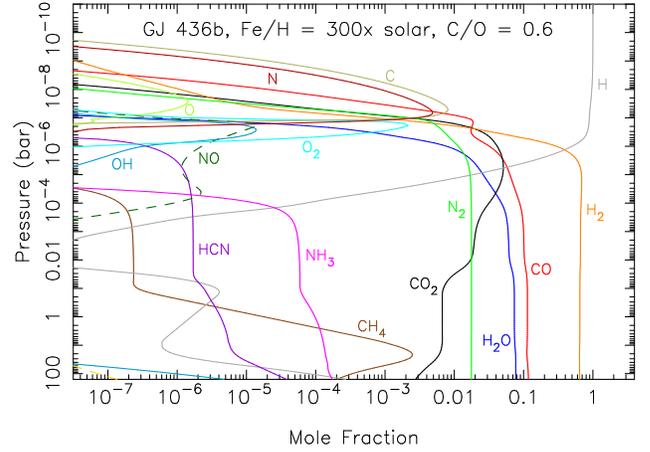}
\caption{Mixing-ratio profiles for several species of interest (as labeled) in our kinetics/transport 
models for GJ 436b, for an assumed atmospheric metallicity 
of 300$\times$ solar in carbon and nitrogen, with the oxygen elemental abundance defined through 
an assumed C/O ratio of 0.6.  The eddy diffusion coefficients are taken to be $K_{zz}$ = 10$^7$ 
$\cmtwo$ $\smone$ at $P$ $\ge$ 10$^{-4}$ bar, with $K_{zz}$ increasing as the 
inverse square root of the pressure at $P$ $<$ 10$^{-4}$ bar.  The thermal profile is 
taken from Fig.~4 of \citet{madhu11gj436b}, but we have shifted the entire profile upwards uniformly 
in an attempt to account for the higher-altitude photosphere with higher metallicities (see text).
A color version of this figure is available in the online journal.\label{fig300x}}
\end{figure}

Although the 1000$\times$ solar metallicity model itself does not provide a good fit to the 
\citet{stevenson10,stevenson12} secondary-eclipse data, and none of the models in Fig.~\ref{figbarmanall} 
fall within the constraints imposed by the retrievals for all the species, the model-data comparisons 
do suggest further avenues to explore with higher-metallicity models.  The larger observed brightness 
temperature at 3.6 $\mu$m in relation to that at 5.8 and 8 $\mu$m in the {\it Spitzer\/} data 
suggests that the atmosphere has a sharper thermal gradient than the one derived from the 1-D 
PHOENIX model, given that the contribution functions at the three wavelengths are not hugely 
separated in terms of the pressures at which they peak in the high-metallicity models.  We 
note that both the dayside-average profiles from the 3-D GCMs of \citet{lewis10} and the 
thermal profiles derived from the retrievals exhibit a larger thermal gradient in the relevant 
photospheric region than the 1000$\times$ solar 1-D PHOENIX models do.  A larger thermal gradient 
could also result in a hotter atmosphere at the CH$_4$-CO quench point, if the quench point occurs 
at a high-enough altitude, pushing the atmosphere towards greater CO dominance at the expense of 
CH$_4$.  That would improve the relative 3.6-to-4.5-$\mu$m flux ratio in comparison with 
observations.  To investigate such a scenario, we simply scale one of the thermal profiles 
derived from the \citet{madhu11gj436b} MCMC analysis upward uniformly in altitude to lower 
pressures and run different disequilibrium chemistry models, assuming different bulk atmospheric 
properties.  The results from one such model are shown in Fig.~\ref{fig300x}.

\begin{figure}
\includegraphics[angle=-90,scale=0.36]{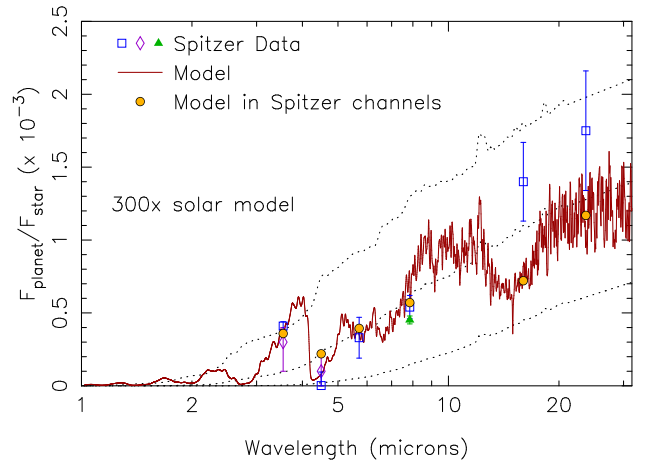}
\caption{Synthetic emission spectra (dark red) for GJ 436b for our disequilibrium model that assumes 
a 300$\times$ solar metallicity atmosphere, with a C/O ratio of 0.6 (see text and Fig.~\ref{fig300x}).
The blue squares represent the {\it Spitzer\/} secondary-eclipse data analyzed by \citet{stevenson10}, 
with an updated upper limit at 4.5-$\mu$m from \citet{stevenson12}, the purple diamonds represent the 
the 3.6 and 4.5-$\mu$m {\it Spitzer\/} analysis of \citet{beaulieu11}, and the green triangle 
represents additional {\it Spitzer\/} eclipse data at 8-$\mu$m as analyzed by \citet{knutson11}.  The 
red circles represent the model fluxes averaged over the {\it Spitzer\/} bandpasses.  The black dotted 
lines represent the planetary emission assuming GJ 436b radiates as a blackbody at a temperature of 500 K 
(lower curve), 800 K (middle curve), or 1100 K (upper curve).  See Fig.~\ref{figspec1x} for further details.
A color version of this figure is available in the online journal.\label{spec300x}}
\end{figure}

For the model presented in this figure, the atmospheric C/H and N/H metallicity are assumed to be 
300$\times$ solar, the C/O ratio above the silicate clouds is assumed to be 0.6 (which actually 
requires a slightly sub-solar initial bulk C/O ratio if $\sim$21\% of the oxygen is sequestered in 
silicates and other condensates at deeper atmospheric levels, as expected by \citealt{lodders10} and 
\citealt{viss10rock}), 
and the eddy diffusion coefficient follows the relation $K_{zz}$ = 10$^7$ $\cmtwo$ $\smone$ at 
pressures $P$ $\ge$ 10$^{-4}$ bar, with $K_{zz}$ increasing with the inverse square root of 
atmospheric pressure at $P$ $<$ 10$^{-4}$ bar.  The thermal profile as shown in the insert of 
Fig.~4 of \citet{madhu11gj436b} has been shifted upward uniformly by $-1.2$ in $\log_{10} (P)$ (i.e., 
pressures have been multiplied by 10$^{-1.2}$) to account for the higher-altitude photosphere 
that will result from the higher metallicity.  The 300$\times$ solar atmospheric metallicity in 
this model is consistent with the interior models for GJ 436b 
\citep{adams08,baraffe08,figueira09,rogers10frame,nettelmann10,miller11}, and the lower value of 
$K_{zz}$ in the lower atmosphere allows CH$_4$ to remain in equilibrium until $\sim$0.1 bar, at which 
point the methane quenches at the desired $<$ 1 ppm abundance.  Judging from the PHOENIX models, the 
thermal profile at photospheric pressures may be hotter in this model than is expected on average 
in the dayside hemisphere of GJ 436b at this metallicity, so the likelihood of this scenario will 
need to be investigated by further 3-D circulation modeling, but given that the observed emission 
will derive preferentially from the hottest regions on the observed disk, the profile does not 
appear to be unreasonably hot.

Figure \ref{fig300x} shows that $\Htwo$ is still the dominant atmospheric constituent (by number) 
in the atmosphere in this 300$\times$ solar metallicity model, although both CO and $\HtwoO$ at mole 
fractions of $\sim$10\% each are encroaching on the $\Htwo$ dominance.  Kinetic production of CO$_2$ 
from $\HtwoO$ and CO occurs in the middle and upper photosphere in this model, allowing the CO$_2$ 
abundance to exceed that of water at high altitudes.  Quenching of ammonia at non-negligible values 
occurs and is responsible for maintaining HCN at greater than ppm levels.  The CH$_4$ abundance 
remains low, and the emission spectrum (Fig.~\ref{spec300x}) is dominated by opacity from $\HtwoO$, 
$\COtwo$, and CO.  This model finally has a low-enough methane abundance that the 3.6-$\mu$m flux from 
the {\it Spitzer\/} secondary-eclipse analysis of \cite{stevenson10} is better reproduced.  The resulting 
3.6-to-4.5-$\mu$m flux ratio is also more in line with observations, although this model, like 
all others including the retrievals, still produces too much flux in the 4.5-$\mu$m band.  This result 
is more a function of the atmosphere being too hot where the contribution function peaks at 4.5 $\mu$m, 
rather than being due to insufficient CO and CO$_2$, as can be readily seen in Figs.~\ref{figbarmanall} \& 
\ref{fig300x}.  In fact, the very large CO$_2$ abundance in this 300$\times$ solar metallicity model 
produces too much absorption in the 16-$\mu$m {\it Spitzer\/} channel, where the contribution function 
peaks at cooler, higher altitudes due to the large CO$_2$ column abundance.  The non-detection of 
the GJ 436b secondary eclipse at 4.5 $\mu$m is difficult to explain by any known physically reasonable 
scenario without adversely affecting the fit at other wavelengths.  In terms of the statistical fit to 
all the \citet{stevenson10,stevenson12} secondary eclipse depths, this model fares better, at 
$\chi^2 /N_{obs}$ = 2.8.  The primary problems are insufficient flux at 3.6, 16, and 24 $\mu$m, 
and too much flux at 4.5 $\mu$m.  

This particular forward model is not unique in terms of fitting the data, and there are numerous models 
with atmospheric metallicities in the couple-hundred to couple-thousand times solar range that 
fit the {\it Spitzer\/} data to within $\chi^2 /N_{obs}$ $\le$ 3 (see, for example the 2000$\times$ 
solar metallicity model we presented in \citealt{richardson13}), but we have not found any 
disequilibrium chemistry models that fit to within $\chi^2 /N_{obs}$ $\le$ 2.  All of these high-metallicity 
models produce too much CO$_2$ to be consistent with the observed flux in the 16-$\mu$m 
channel.  Models with high metallicities in the 100--10,000$\times$ solar range also end up with a 
large water abundance, which then requires a hot photosphere to explain the 5.8, 8.0, and 24-$\mu$m 
fluxes, at which point there is too much emission at 4.5 $\mu$m.  Slightly cooler models than the one 
shown in Figs.~\ref{fig300x} \& \ref{spec300x} tend to fit better at 4.5 $\mu$m, but then the fit tends 
to be worse at 3.6 and 24 $\mu$m.  Hotter models fit the 24-$\mu$m flux better, but then the fit at 4.5, 
5.8, and 8.0 $\mu$m is worse.  Going to a higher C/O ratio will help remove excess CO$_2$ to improve 
the fit at 16 $\mu$m, but then CH$_4$ becomes too abundant to explain the 3.6-$\mu$m flux.  Lower 
C/O ratios help keep the methane abundance low, but then the problems with excess $\HtwoO$ and 
$\COtwo$ are exacerbated.  Therefore, while these high-metallicity models are promising in their 
ability to qualitatively reproduce several aspects of the observed secondary-eclipse observations, 
several quantitative problems remain.  The solutions derived from the retrieval methods 
(section~\ref{sectretrieve} and \citealt{madhu11gj436b}) provide a better overall fit to the data, 
but problems with the fit still remain at 4.5 and/or 16 $\mu$m; more importantly, these solutions 
have problems with plausibility in that it is difficult to theoretically explain how such abundances 
could be obtained chemically or physically in an atmosphere with any reasonable bulk properties.  
Therefore, despite the less-than-perfect fit to the existing GJ 436b secondary-eclipse data, such 
high-metallicity models should not be dismissed out of hand.

\begin{figure}
\includegraphics[scale=0.36]{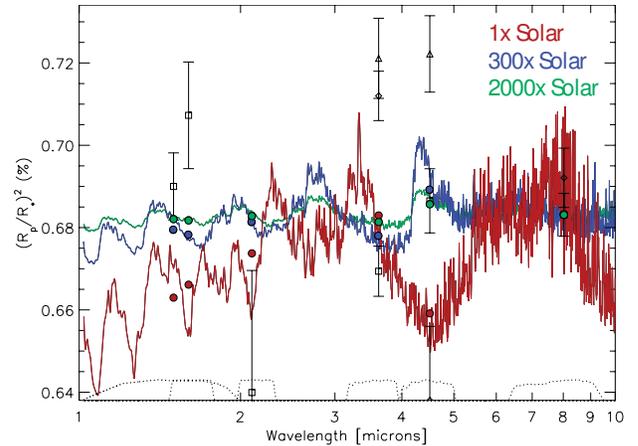}
\caption{Synthetic transmission spectra for GJ 436b (in terms of the apparent transit depth, i.e., the 
square of the ratio of the planetary radius to the stellar radius) for our disequilibrium models that 
assume a 1$\times$ (red), 300$\times$ (blue), and 2000$\times$ (green) solar metallicity.  Scattering 
from molecules or hazes is not included in the calculations.  Data points 
from various sources are shown in black, with associated error bars.  In order of increasing 
wavelength, the squares are from the 1.1--1.9 $\mu$m HST/NICMOS analysis of \citet{pont09} plotted at 
1.5 $\mu$m; the ground-based $H$-band data of \citet{alonso08} plotted at 1.6 $\mu$m; the ground-based 
$K_s$ band data of \citet{caceres09} plotted at at 2.1 $\mu$m, as discussed by \citet{knutson11}; and 
the {\it Spitzer}/IRAC analysis of \citet{knutson11} plotted at 
3.6, 4.5, and 8 $\mu$m.  The diamonds are from the {\it Spitzer\/} analysis of \citet{beaulieu11} at 
3.6, 4.5, and 8 $\mu$m.  The triangles are the {\it Spitzer\/} transits at 3.6 and 4.5 $\mu$m that 
\citet{knutson11} suggest are influenced by the occultation of star spots or other regions of non-uniform 
brightness on the star.  The black dotted curves at the bottom show the response functions for 
the filters and/or detector channels used in the observations.  The colored circles represent the 
corresponding model transit depths averaged over the appropriate filters/channels.  
A color version of this figure is available in the online journal.\label{transit}}
\end{figure}

The higher the atmospheric metallicity, the higher the mean molecular weight, and the smaller the 
atmospheric scale height for any given temperature and gravity profile.  High-metallicity atmospheres 
will then have flatter transmission spectra \citep[e.g.,][]{fortney13}.  Examples of transmission 
spectra from some of our GJ 436b models are shown in Fig.~\ref{transit}, in comparison with 
observations.  The relatively flat spectrum observed in HST/NICMOS data \citep{pont09,gibson11} 
seems more consistent with the high-metallicity models, and such models also seem to compare well to the 
results from the \citet{knutson11} {\it Spitzer\/} analysis (see the open squares at 3.6, 4.5, and 8 $\mu$m 
in Fig.~\ref{transit}).  However, the high-metallicity models do not compare well with the \citet{beaulieu11} 
{\it Spitzer\/} transit analysis (see the open diamonds in Fig.~\ref{transit}) or the $K_s$-band transit 
data of \citet{caceres09}, and in fact none of the models provides a reasonable fit to all the available 
transit data for GJ 436b.  Given the observed variability in the apparent transit depths with time from 
different observations \citep[e.g.,][]{knutson11} and the lack of agreement between transit results 
analyzed with different procedures \citep[e.g.,][]{pont09,gibson11,knutson11,beaulieu11}, the poor fits 
of these models to the transit data are not particularly surprising.  Obtaining simultaneous spectral 
observations at multiple wavelengths should help resolve some of these issues and help distinguish between 
competing models.

Note that we have not adjusted for potential differences in temperatures or chemical abundances 
between the terminator and dayside atmospheres when performing these calculations.  The dayside 
atmosphere observed at secondary eclipse is likely hotter than the limb atmosphere at the terminators 
probed in the planetary transit \citep[see][]{lewis10}.  The non-negligible eccentricity of GJ 436b 
will also lead to variations in temperature over the orbit, which is captured in the GCMs of 
\citet{lewis10}.  It is likely that horizontal quenching will help homogenize the atmospheric 
abundances with respect to longitude due to strong zonal winds or rapid thermal changes over the 
orbit \citep{coop06,iro10,visscher12,agundez12}, but without more detailed multi-dimensional and temporal 
modeling, it is difficult to predict the exact terminator composition.  Regardless of the exact 
molecular composition, however, the high-metallicity models will have a higher atmospheric mean 
molecular mass, and a flatter overall transit spectrum.

\begin{figure*}
\begin{center}
\includegraphics[scale=0.7]{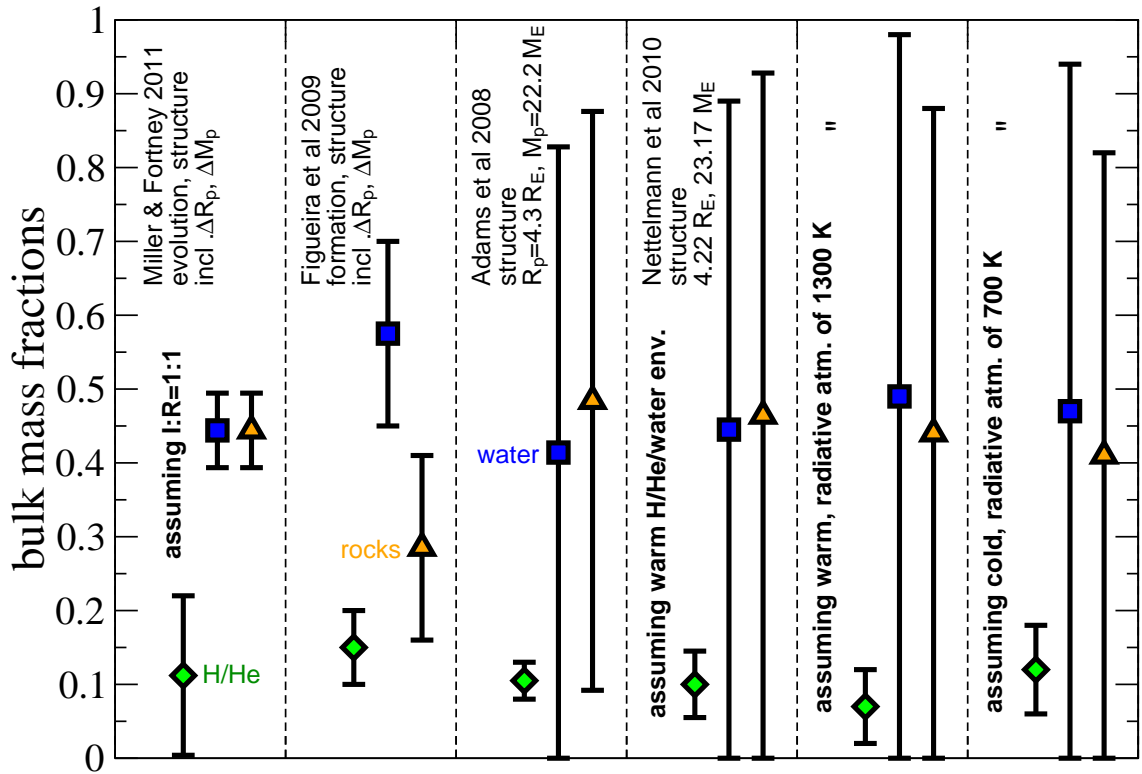}
\caption{Constraints on the bulk composition of GJ 436b as determined from models of its structure, 
evolution, and formation.  Possible compositions are shown for six different model assumptions 
and/or imposed constraints.  For each, the sum over the displayed mass fractions of H/He (green 
diamonds), water (blue squares), and ``rock'' (includes silicates plus iron; orange triangles) 
equals 1.  All calculations take into account the planet's mass $M_p$ and radius $R_p$ as one set 
of constraints; the two leftmost models consider uncertainties in $M_p$ and $R_p$, while the four 
on the right adopt specific $M_p$ and $R_p$ values from \citet{torres08}.  From left to right, the 
derived constraints are based on (i) the thermal and structural evolution models of \citet{miller11} 
that assume an ice:rock ratio of 1:1 and that impose constraints based on the age of the GJ 436 
system; (ii) the formation, disk-planet evolution (including migration), and structural models of 
\citet{figueira09} that compute various formation paths consistent with $M_p$ and $R_p$; (iii) the 
structure models of \citet{adams08} that consider precise values for $M_p$, $R_p$, and atmospheric 
temperatures; (iv) the structural models of \citet{nettelmann10} that assume that water is mixed 
into the outermost H/He envelope; (v) structural models based on \citet{nettelmann10} that assume 
that a pure H/He layer resides above a water layer with a warm (1300 K) deep isothermal radiative 
region; (vi) a model similar to the one on the immediate left, except the isothermal radiative region 
is cool (700 K) and the planet is old ($>$ 10 Gyr).  Note that the tightest constraints on the 
composition are imposed by the formation models, that including the uncertainties in $M_p$ and $R_p$ 
enhance the uncertainty in the H/He mass fraction by a factor of $\sim$2, and that assumptions about 
atmospheric temperatures affect the derived H/He mass fraction.  
A color version of this figure is available in the online journal.\label{fignadine}}
\end{center}
\end{figure*}

\section{GJ 436\lowercase{b} Compositional Constraints from Structure and Evolution Models}

As mentioned earlier, a relatively high metallicity for GJ 436b is also indicated by the planet's 
observed radius in relation to its mass.  Figure~\ref{fignadine} shows some of the compositional 
constraints imposed by models of the planet's interior structure, formation, and evolution 
\citep[e.g.,][]{adams08,baraffe08,figueira09,rogers10frame,nettelmann10,miller11}.  The 
amount of H/He required to explain the mass and radius of GJ 436b depends on several other 
assumptions in the models, such as the relative abundance of ices (i.e., volatile heavy 
elements) versus rocks (i.e., refractory heavy elements), the age and evolutionary history 
of the planet, the presence/absence of a central core, the degree of mixing of elements within 
the planet, and the thermal structure within the planet.  Not much H/He is needed to explain the 
observed radius if the planet is very hot (i.e., young, high intrinsic interior $T_{int}$ or tidal
heating, early onset of convective interior, high-temperature isothermal radiative region), if the 
heavy elements in the planet have low atomic weight (i.e., ices, not rocks), and/or if the heavy 
elements are confined to deep levels and are not mixed throughout the atmospheric envelope.  
Conversely, more H/He is needed to explain the radius if the planet is old and cool, contains 
more rocky than icy elements, and has heavy elements mixed throughout the atmospheric envelope.  
Parameters such as the ice-to-rock ratio of GJ 436b are not well constrained from structure 
and evolution models alone, and the planet's age, evolutionary history, and atmospheric and 
interior thermal structure are also uncertain.  As is clear from Fig.~\ref{fignadine}, the 
formation models of \citet{figueira09} provide the tightest constraints on the bulk composition.   
When the uncertainties in GJ 436b's mass, radius, and age are taken into account 
\citep[e.g.,][]{miller11}, the final uncertainty in the possible H/He mass fraction for 
GJ 436b is significantly increased.  The assumed temperature of the isothermal radiative 
portion of the atmosphere somewhat influences the inferred H/He mass fraction \citep{nettelmann10}.

Accounting for the various uncertainties and adopting reasonable assumptions about the atmosphere 
and interior, most of these models predict H/He mass fractions of $\sim$3--22\% for GJ 436b, 
corresponding to metallicities of $\sim$230--2000$\times$ solar.  \citet{nettelmann10} demonstrate that 
even lower H/He mass fractions (higher metallicities) are possible if the planet contains few rocky 
elements (only ices), if the ices are confined to the core, and if the pure H/He outer envelope 
is hot as a result of a thin radiative region.  A hot isothermal radiative region is possible for 
GJ 436b (see \citealt{spiegel10} and our Fig.~\ref{figtemp} above); however, the other assumptions 
are less secure, and given that the secondary eclipse data show evidence for molecular absorbers 
rather than a blackbody atmosphere, it seems likely that at least some heavy elements are mixed 
into the outer atmospheric envelope.  Therefore, the $\sim$230--2000$\times$ solar models seem most 
probable for GJ 436b, and our high-metallicity disequilibrium models discussed in section 
\ref{sectdisequil} appear reasonable in terms of the interior and evolution models for GJ 436b.

\section{Conclusions}

Moderate-to-high metallicity models ($\sim$230-2000$\times$ solar) for GJ 436b are appealing 
in that they provide a natural explanation for the apparent CO-rich, CH$_4$-poor nature of 
GJ 436b's atmosphere \citep[see][]{stevenson10,madhu11gj436b} and are consistent with the 
heavy-element enrichment as inferred from the planet's mass and radius from interior modeling 
\citep{adams08,baraffe08,figueira09,rogers10frame,nettelmann10}.  Although such high-metallicity 
models are not without problems in terms of explaining all the transit and secondary-eclipse 
data (see section~\ref{sectdisequil}), they do have one significant advantage over the solutions 
obtained to date from retrieval methods (e.g., \citealt{madhu11gj436b} and section~\ref{sectretrieve} 
above) --- the advantage of physical and chemical plausibility.  The relative abundances of CO, 
$\HtwoO$, $\CHfour$, and $\COtwo$ derived for GJ 436b from the retrieval methods do not appear 
to be achievable from either equilibrium or disequilibrium chemistry in a GJ 436b atmosphere 
with any plausible imagined bulk properties (i.e., metallicity, C/O ratio, temperature).  Either 
our theoretical understanding is incomplete, and we are missing some key disequilibrium mechanisms 
that convert $\CHfour$ and $\HtwoO$ permanently to CO in GJ 436b's dayside atmosphere, or the 
retrieved solutions simply reflect some of the pitfalls that arise when applying retrieval 
methods to sparse data sets with large systematic uncertainties \citep[e.g.,][]{line13}.  From the latter 
standpoint, it would be useful if the error bars provided for transit and eclipse observations accurately 
reflected systematic uncertainties inherent with the instruments, along with the more commonly 
reported statistical uncertainties.  Failing that, instrument systematics could be considered when 
performing the retrievals.

To make further progress on characterizing the atmosphere of GJ 436b (and other hot Neptunes), we 
would need additional transit and eclipse data, preferably from a space-based instrument that can 
deliver spectra at multiple wavelengths simultaneously.  Given the observed variability in 
the GJ 436b transit depths from one transit to the next \citep{knutson11}, which might be related to 
the occultation of star spots, obtaining simultaneous wavelength information is desirable for 
transit observations.  
If hot Netpune exoplanets do indeed have higher metallicities than is typical for hot Jupiters, as we 
suggest, then the transmission spectra from hot Neptunes would be flatter and more featureless in 
general, even in the absence of photospheric clouds.  Figure~\ref{transit} shows that transit observations 
with moderate spectral resolution in the 1--5 $\mu$m region should help distinguish between the low- or 
high-metallicity scenarios, with the relative transit depth at $\sim$3.5 $\mu$m versus 4.5 $\mu$m being 
particularly diagnostic.  Emission spectra from secondary eclipse observations have some advantages over 
transit observations in terms of characterizing atmospheres, despite the well-known problems with 
degeneracies between temperatures and abundances \citep[e.g.,][]{burrows10}, because more information can 
be derived with respect to atmospheric temperatures, the emission measurements tend to be less sensitive 
to vertically-thin scattering clouds and hazes, and the planetary flux can compete with the stellar flux 
at mid-IR wavelengths where many molecules have diagnostic features.  In that regard, having a space-based 
infrared telescope that extends to $\sim$16 microns, like EChO \citep{tinetti12}, could be critical for 
correctly diagnosing atmospheric properties \citep[e.g.,][]{tinetti12,barstow13}.

Regardless of whether GJ 436b itself has a high-metallicity atmosphere, radial-velocity and transit 
surveys show that Neptune-class exoplanets are exceedingly common in the galaxy 
\citep{howard10,howard12,mayor11,borucki11,batalha13}, and based on our current views of exoplanet 
formation (see section 3.1), we would expect many of these planets to have atmospheric metallicities 
that are orders of magnitude in excess of solar.  There is some evidence for that trend in the existing 
population of exoplanets for which we have derivations of both mass and radius \citep[e.g.,][]{miller11,weiss13}.  
As is shown in Figs.~\ref{figpie}, ~\ref{figtempvsmetal}, \& \ref{figtempvsratio}, hot Neptunes 
could have very diverse atmospheric compositions depending on atmospheric temperatures and bulk atmospheric 
properties such as metallicity and C/O ratio.  The higher-metallicity 
atmospheres described in this paper have no analogs from within our own solar system, but one 
can imagine a continuum of hot Neptunes that should exist in the exoplanet population, with atmospheres 
ranging from moderate-to-low metallicity, hydrogen-rich, Neptune-like compositions to high-metallicity, 
hydrogen-poor, super-Venus-like compositions, along with more exotic CO-dominated planets and 
$\HtwoO$-dominated ``water worlds''.  As the statistics regarding exoplanet properties continues to 
grow, so does our amazement at the diversity of these worlds beyond the narrow confines of our solar system.



\acknowledgments

We thank Travis Barman for calculating and providing 1-D thermal profiles for high-metallicity 
GJ 436b atmospheres, Nikole Lewis for providing her dayside-average GCM thermal profiles for 
GJ 436b, and Heather Knutson for useful conversations.
This work was supported by the NASA Planetary Atmospheres Program grant number NNX11AD64G.





\clearpage


\begin{thebibliography}{}
\bibitem[Adams et al.(2008)]{adams08} Adams, E. R.,
   Seager, S., \& Elkins-Tanton, L. 2008, \apj, 673, 1160
\bibitem[Ag\'undez et al.(2012)]{agundez12} Ag\'undez, M.,
   Venot, O., Iro, N., et al. 2012, A\&A, 548, A73
\bibitem[Alibert et al.(2005)]{alibert05} Alibert, Y.,
   Mordasini, C., Benz, W., \& Winisdoerffer, C. 2005, A\&A, 434, 343
\bibitem[Alibert et al.(2006)]{alibert06} Alibert, Y.,
   Baraffe, I., Benz, W., et al. 2006, A\&A, 455, L25
\bibitem[Allard et al.(2001)]{allard01} Allard, F.,
   Hauschildt, P. H., Alexander, D. R., Tamanai, A., \& Schweitzer, A. 
   2001, \apj, 556, 357
\bibitem[Allen et al.(1981)]{allen81} Allen, M.,
   Yung, Y. L., \& Waters, J. W. 1981, J. Geophys. Res., 86, 3617
\bibitem[Alonso et al.(2008)]{alonso08} Alonso, R.,
   Barbieri, M., Rabus, M., et al. 2008, A\&A, 487, L5
\bibitem[Baines et al.(1995)]{baines95} Baines, K. H., 
   Mickelson, M. E., Larson, L. E., \& Ferguson, D. W. 1995, Icarus, 114, 328
\bibitem[Ballard et al.(2010)]{ballard10} Ballard, S.,
   Christiansen, J. L., Charbonneau, D., et al. 2010, \apj, 716, 1047
\bibitem[Baraffe et al.(2008)]{baraffe08} Baraffe, I.,
   Chabrier, G., \& Barman, T. 2008, A\&A, 482, 315
\bibitem[Barman et al.(2001)]{barman01} Barman, T. S.,
   Hauschildt, P. H., \& Allard, F. 2001, \apj, 556, 885
\bibitem[Barman et al.(2005)]{barman05} Barman, T. S.,
   Hauschildt, P. H., \& Allard, F. 2005, \apj, 632, 1132
\bibitem[Barstow et al.(2013)]{barstow13} Barstow, J. K.,
   Aigrain, S., Irwin, P. G. J., et al. 2013, MNRAS, 430, 1188
\bibitem[Batalha et al.(2013)]{batalha13} Batalha, N. M., 
   Rowe, J. F., Bryson, S. T., et al. 2013, \apjs, 204, 24
\bibitem[Bean et al.(2006)]{bean06} Bean, J. L.,
   Benedict, G. F., \& Endl, M. 2006, \apj, 653, L65
\bibitem[Bean et al.(2008)]{bean08} Bean, J. L.,
   Benedict, G. F., Charbonneau, D., et al. 2008, A\&A, 486, 1039
\bibitem[Beaulieu et al.(2011)]{beaulieu11} Beaulieu, J.-P., 
   Tinetti, G., Kipping, D. M., et al. 2011, \apj, 731, 16
\bibitem[Benneke \& Seager(2012)]{benneke12} Benneke, B.,
   \& Seager, S., 2012, \apj, 753, 100
\bibitem[B\'ezard et al.(2002)]{bezard02} B\'ezard, B.,
   Lellouch, E., Strobel, D., Maillard, J.-P., \& Drossart, P. 2002, 
   Icarus, 159, 95
\bibitem[Bodenheimer \& Pollack(1986)]{bodenheimer86} Bodenheimer, P. H.,
   \& Pollack, J. B., 1986, Icarus, 67, 391
\bibitem[Borucki et al.(2011)]{borucki11} Borucki, W. J.,
   Koch, D. G., Basri, G., et al. 2011, \apj, 736, 19
\bibitem[Burrows \& Orton(2010)]{burrows10} Burrows, A.,
   \& Orton, G. 2010, in Exoplanets, ed. S. Seager (Tucson: Univ. Arizona Press), 419
\bibitem[Butler et al.(2004)]{butler04} Butler, R. P.,
   Vogt, S. S., Marcy, G. W., et al. 2004, \apj, 617, 580
\bibitem[C\'aceres et al.(2009)]{caceres09} C\'aceres, C., 
   Ivanov, V. D., Minniti, D., et al. 2009, A\&A, 507, 481
\bibitem[Cooper \& Showman(2006)]{coop06} Cooper, C. S., \&
   Showman, A. P., 2006, \apj, 649, 1048
\bibitem[Coughlin et al.(2008)]{coughlin08} Coughlin, J. L.,
   Stringfellow, G. S., Becker, A. C., et al. 2008, \apj, 689, L149
\bibitem[Deming et al.(2007)]{deming07} Deming, D., 
   Harrington, J., Laughlin, G., et al. 2007, \apj, 667, L199
\bibitem[Demory et al.(2007)]{demory07} Demory, B.-O., 
   Gillon, M., Barman, T., et al. 2007, A\&A, 475, 1125
\bibitem[Ehrenreich et al.(2011)]{ehrenreich11} Ehrenreich, D.,
   Lecavelier des Etangs, A., \& Delfosse, X. 2011, A\&A, 529, A80
\bibitem[Elkins-Tanton \& Seager(2008)]{elkins08} Elkins-Tanton, L. T.,
   \& Seager, S. 2008, \apj, 685, 1237
\bibitem[Fegley \& Lodders(1994)]{fegley94} Fegley, B., Jr., 
   \& Lodders, K. 1994, Icarus, 110, 117
\bibitem[Fegley et al.(1991)]{fegley91} Fegley, B., Jr.,
   Gautier, D., Owen, T., \& Prinn, R. G. 1991, in Uranus, 
   ed. J. T. Bergstrahl, E. D. Miner, M. Shapley Matthews, (Tucson:Univ.~Arizona Press), 147
\bibitem[Figueira et al.(2009)]{figueira09} Figueira, P.,
   Pont, F., Mordasini, C., et al. 2009, A\&A, 493, 671
\bibitem[Fortney et al.(2006)]{fort06hd149} Fortney, J. J., 
   Saumon, D., Marley, M. S., Lodders, K., \& Freedman, R. S. 2006, 
   \apj, 642, 495
\bibitem[Fortney et al.(2007)]{fortney07rad} Fortney, J. J.,
   Marley, M. S., \& Barnes, J. W. 2007, \apj, 659, 1661
\bibitem[Fortney et al.(2013)]{fortney13} Fortney, J. J.,
   Mordasini, C., Nettelmann, N., Kempton, E., \& Zahnle, K. 2013, \apj, submitted
\bibitem[Fortney \& Nettelmann(2010)]{fortnet10} Fortney, J. J.,
   \& Nettelmann, N. 2010, Space Sci. Rev., 152, 423
\bibitem[Fouchet et al.(2009)]{fouchet09} Fouchet, T.,
   Moses, J. I., \& Conrath, B. J. 2009, in Saturn from Cassini-Huygens,
   ed.~M. K. Dougherty, L. W. Esposito, \& S. M. Krimigis (Dordrecht:Springer), 83
\bibitem[France et al.(2013)]{france13} France, K.,
   Froning, C. S., Linsky, J. L., et al. 2013, \apj, 763, 149
\bibitem[Fressin et al.(2013)]{fressin13} Fressin, F.,
   Torres, G., Charbonneau, D., et al. 2013, \apj, 766, 81
\bibitem[Gaidos(2012)]{gaidos12} Gaidos, E.
   2012, abstract presented at Comparative Climatology of Terrestrial Planets, 
   25-28 June 2012 in Boulder, CO, USA, 8037
\bibitem[Garc\'{\i}a Mu\~noz(2007)]{garcia07} Garc\'{\i}a Mu\~noz, A.
   2007, Planet. Space Sci., 55, 1426
\bibitem[Gautier et al.(1995)]{gautier95} Gautier, D.,
   Conrath, B. J., Owen, T., de Pater, I., \& Atreya, S. K. 1995, in Neptune and Triton, 
   ed. D. P. Cruikshank, (Tucson:Univ.~Arizona Press), 547
\bibitem[Gibson et al.(2011)]{gibson11} Gibson, N. P.,
   Pont, F., \& Aigrain, S. 2011, MNRAS, 411, 2199
\bibitem[Gillon et al.(2007a)]{gillon07a} Gillon, M., 
   Demory, B.-O., Barman, T., et al. 2007a, A\&A, 471, L51
\bibitem[Gillon et al.(2007b)]{gillon07b} Gillon, M., 
   Pont, F., Demory, B.-O., et al. 2007b, A\&A, 472, L13
\bibitem[Gordon \& McBride(1994)]{gordon94} Gordon, S.,
   \& McBride, B. J. 1994, NASA Reference Publ.~1311
\bibitem[Guillot(2010)]{guillot10} Guillot, T., 2010, A\&A, 527, A27
\bibitem[Hansen \& Murray(2012)]{hansen12} Hansen, B., 
   \& Murray, N. 2012, \apj, 751, 158
\bibitem[Hauschildt et al.(1999)]{hauschildt99} Hauschildt, P. H., 
   \& Baron, E. 1999, J. Comput. Appl. Math., 109, 41
\bibitem[Hawley et al.(1996)]{hawley96} Hawley, S. L.,
   Gizis, J. E., \& Reid, I. N. 1996, AJ, 112, 2799
\bibitem[Heap \& Lindler(2010)]{heap10} Heap, S.,
   \& Lindler, D. 2010, http://archive.stsci.edu/pub/hlsp/stisngsl/
\bibitem[Howard et al.(2010)]{howard10} Howard, A. W.,
   Marcy, G. W., Johnson, J. A., et al. 2010, Science, 330, 653
\bibitem[Howard et al.(2012)]{howard12} Howard, A. W.,
   Marcy, G. W., Bryson, S. T., et al. 2012, \apjs, 201, 15
\bibitem[Hu et al.(2012)]{hu12} Hu, R.,
   Seager, S., \& Bains, W. 2012, \apj, 761, 166
\bibitem[Hubbard et al.(1995)]{hubbard95} Hubbard, W. B.,
   Podolak, M., \& Stevenson, D. J. 1995, in Neptune and Triton, 
   ed. D. P. Cruikshank, (Tucson:Univ.~Arizona Press), 109
\bibitem[Iro \& Deming(2010)]{iro10} Iro, N, 
   \& Deming, L. D. 2010, \apj, 712, 218
\bibitem[Jenkins et al.(2009)]{jenkins09} Jenkins, J. S.,
   Ramsey, L. W., Jones, H. R. A., et al. 2009, \apj, 704, 975
\bibitem[Karkoschka \& Tomasko(2009)]{karkoschka09} Karkoschka, E.,
   \& Tomasko, M. 2009, Icarus, 202, 287
\bibitem[Karkoschka \& Tomasko(2011)]{karkoschka11} Karkoschka, E.,
   \& Tomasko, M. 2011, Icarus, 211, 780
\bibitem[Kirkpatrick et al.(1991)]{kirkpatrick91} Kirkpatrick, J. D.,
   Henry, T. J., \& McCarthy, D. W., Jr. 1991, \apjs, 77, 417
\bibitem[Kite et al.(2009)]{kite09} Kite, E. S.,
   Manga, M., \& Gaidos, E. 2009, \apj, 700, 1732
\bibitem[Knutson et al.(2010)]{knutson10} Knutson, H. A., 
   Howard, A. W., \& Isaacson, H. 2010, \apj, 720, 1569
\bibitem[Knutson et al.(2011)]{knutson11} Knutson, H. A., 
   Madhusudhan, N., Cowan, N. B., et al. 2011, \apj, 735, 27
\bibitem[Kopparapu et al.(2012)]{koppa12} Kopparapu, R. K.,
   Kasting, J. F., \& Zahnle, K. J. 2012, \apj, 745, 77
\bibitem[Kuchner \& Seager(2005)]{kuchner05} Kuchner, M. J.,
   \& Seager, S. 2005, arXiv:astro-ph/0504214
\bibitem[Lee et al.(2012)]{lee12} Lee, J.-M., 
   Fletcher, L.N., \& Irwin. P.G.J. 2011, MNRAS, 420, 170
\bibitem[Lewis \& Fegley(1984)]{lewis84} Lewis, J. S.,
   \& Fegley, B., Jr. 1984, \ssr, 39, 163
\bibitem[Lewis et al.(2010)]{lewis10} Lewis, N. K.,
   Showman, A. P., Fortney, J. J., et al. 2010, \apj, 720, 344
\bibitem[Liang et al.(2003)]{liang03} Liang, M.-C.,
   Parkinson, C. D., Lee, A. Y. T., Yung, Y. L., \& Seager, S. 
   2003, \apj, 596, L247
\bibitem[Liang et al.(2004)]{liang04} Liang, M.-C.,
   Seager, S., Parkinson, C. D., Lee, A. Y. T., \& Yung, Y. L.
   2004, \apj, 605, L61
\bibitem[Lindzen(1981)]{lindzen81} Lindzen, R. S. 1981,
   J. Geophys. Res., 86, 9707
\bibitem[Line et al.(2010)]{line10} Line, M. R.,
   Liang, M. C., \& Yung, Y. L.  2010, \apj, 717, 496
\bibitem[Line et al.(2011)]{line11} Line, M. R.,
   Vasisht, G., Chen, P., Angerhausen, D., \& Yung, Y. L. 2011, \apj, 738, 32
\bibitem[Line et al.(2012)]{line12} Line, M. R.,
   Zhang, X., Vasisht, G., et al. 2012, \apj, 749, 93
\bibitem[Line et al.(2013)]{line13} Line, M. R.,
   Wolf, A. S., Zhang, X., et al. 2013, arXiv: 1304.5561
\bibitem[Lissauer \& Stevenson(2007)]{lissauer07} Lissauer, J. J.,
   \& Stevenson, D. J.  2007, in Protostars and Planets V, 
   ed.~B.~Reipurth, D.~Jewitt, \& K. Keil (Tucson, AZ: Univ.~Arizona Press), 591
\bibitem[Lodders(2004)]{lodders04} Lodders, K.
   2004, \apj, 611, 587
\bibitem[Lodders(2010)]{lodders10} Lodders, K. 2010, 
   in Formation and Evolution of Exoplanets, ed.~R. Barnes (Berlin: Wiley), 
   157
\bibitem[Lodders et al.(2009)]{lodders09} Lodders, K.,
   Palme, H., \& Gail, H.-P. 2009, in The Solar System, Landolt-B\"ornstein, New Series, Vol. VI/4B, 
   ed. J. E. Tr\"umper, (Berlin:Springer-Verlag), 560
\bibitem[Lodders \& Fegley(1997)]{lodders97} Lodders, K.,
   \& Fegley, B., Jr. 1997, AIP Conf. Proc., 402, 391
\bibitem[Lodders \& Fegley(1994)]{lodders94} Lodders, K.,
   \& Fegley, B., Jr. 1994, Icarus, 112, 368
\bibitem[Lodders \& Fegley(2002)]{lodders02} Lodders, K.,
   \& Fegley, B., Jr. 2002, Icarus, 155, 393
\bibitem[Luszcz-Cook \& de Pater(2013)]{luszcz13} Luszcz-Cook, S. H.,
   \& de Pater, I. 2013, Icarus, 222, 379
\bibitem[Madhusudhan(2012)]{madhu12} Madhusudhan, N.
   2012, \apj, 758, 36
\bibitem[Madhusudhan et al.(2011a)]{madhu11wasp12b} Madhusudhan, N.,
   Harrington, J., Stevenson, K. B., et al. 2011a, Nature, 469, 64
\bibitem[Madhusudhan et al.(2011b)]{madhu11carbrich} Madhusudhan, N.,
   Mousis, O., Johnson, T.V., \& Lunine, J.I. 2011b, \apj, 743, 191
\bibitem[Madhusudhan \& Seager(2009)]{madhu09} Madhusudhan, N.,
   \& Seager, S. 2009, \apj, 707, 24
\bibitem[Madhusudhan \& Seager(2011)]{madhu11gj436b} Madhusudhan, N.,
   \& Seager, S. 2011, \apj, 729, 41
\bibitem[Maness et al.(2007)]{maness07} Maness, H. L.,
   Marcy, G. W., Ford, E. B., et al. 2007, \pasp, 119, 90
\bibitem[Marley et al.(2007)]{marley07} Marley, M. S.,
   Fortney, J. J., Hubickyj, O., Bodenheimer, P, \& Lissauer, J. J. 2007, \apj, 655, 541
\bibitem[Mayor et al.(2011)]{mayor11} Mayor, M.,
   Marmier, M., Lovis, C. et al. 2011, arXiv: 1109.2497
\bibitem[Miguel et al.(2011)]{miguel11} Miguel, Y.,
   Kaltenegger, L., Fegley, B., Jr., \& Schaefer, L. 2011, \apj, 742, L19
\bibitem[Miller et al.(2009)]{miller09} Miller, J. A.,
   Klippenstein, S. J., Robertson, S. H., Pilling, M. J., \& Green, 
   N. J. B. 2009, Phys. Chem. Chem. Phys., 11, 1128
\bibitem[Miller \& Fortney(2011)]{miller11} Miller, N.,
   \& Fortney, J. J. 2011, \apj, 736, L29
\bibitem[Miller-Ricci \& Fortney(2010)]{miller-ricci10} Miller-Ricci, E.,
   \& Fortney, J. J. 2010, \apj, 716, L74
\bibitem[Miller-Ricci Kempton et al.(2012)]{miller-ricci12} Miller-Ricci Kempton, E.,
   Zahnle, K., \& Fortney, J. 2012, \apj, 745, 3
\bibitem[Mizuno et al.(1978)]{mizuno78} Mizuno, H.,
   Nakazawa, K., \& Hayashi, C. 1978, Prog. Theor. Phys., 60, 699
\bibitem[Mordasini et al.(2012a)]{mordasini12a} Mordasini, C.,
   Alibert, Y., Georgy, C., et al., 2012a, A\&A, 547, A112
\bibitem[Mordasini et al.(2012b)]{mordasini12b} Mordasini, C.,
   Alibert, Y., Klahr, H., \& Henning, T. 2012b, A\&A, 547, A111
\bibitem[Moses(2013)]{moses13rev} Moses, J. I. (2013), Phil.~Trans.~Roy.~Soc.~A, 
   submitted
\bibitem[Moses et al.(2004)]{moses04} Moses, J. I.,
   Fouchet, T., Yelle, R. V., et al. 2004, in Jupiter: 
   The Planet, Satellites \& Magnetosphere, ed.~F. Bagenal, T. E. Dowling, 
   \& W. B. McKinnon (Cambridge: Cambridge Univ. Press), 129
\bibitem[Moses et al.(2010)]{moses10} Moses, J. I.,
   Visscher, C., Keane, T. C., \& Sperier, A. 2010, Faraday Disc., 147, 103
\bibitem[Moses et al.(2011)]{moses11} Moses, J. I.,
   Visscher, C., Fortney, J. J., et al. 2011, \apj, 737, 15
\bibitem[Moses et al.(2013)]{moses13} Moses, J. I.,
   Madhusudhan, N., Visscher, C., Freedman, R. S. 2013, \apj, 763, 25
\bibitem[Mousis et al.(2011)]{mousis11} Mousis, O.,
   Lunine, J. I., Petit, J.-M., et al. 2011, \apj, 727, 77
\bibitem[Mousis et al.(2012)]{mousis12} Mousis, O.,
   Lunine, J. I., Madhusudhan, N., \& Johnson, T. V. 2012, \apj, 751, L7
\bibitem[Nettelmann et al.(2010)]{nettelmann10} Nettelmann, N.,
   Kramm, U., Redmer, R., \& Neuh\"auser, R. 2010, A\&A, 523, A26
\bibitem[Nettelmann et al.(2011)]{nettelmann11} Nettelmann, N.,
   Fortney, J. J., Kramm, U., \& Redmer, R. 2011, \apj, 733, 2
\bibitem[Nettelmann et al.(2013)]{nettelmann13} Nettelmann, N.,
   Helled, R., Fortney, J. J., \& Redmer, R. 2013, Planet. Space Sci., 77, 143
\bibitem[\"Oberg et al.(2011)]{oberg11} \"Oberg, K. I.,
   Murray-Clay, R., \& Bergin, E. A. 2011, \apj, 743, L16
\bibitem[Parmentier et al.(2013)]{parmentier13} Parmentier, V.,
   Showman, A. P., \& Lian, Y. 2013, arXiv: 1301.4522
\bibitem[Podolak et al.(1995)]{podolak95} Podolak, M.,
   Weizman, A., \& Marley, M. 1995, Planet. Space Sci., 43, 1517
\bibitem[Pollack et al.(1996)]{pollack96} Pollack, J. B.,
   Hubickyj, O., Bodenheimer, P., et al. 1996, Icarus, 124, 62
\bibitem[Pont et al.(2009)]{pont09} Pont, F.,
   Gilliland, R. L., Knutson, H., Holman, M., \& Charbonneau, D. 2009, MNRAS, 393, L6
\bibitem[Prinn \& Barshay(1977)]{prinn77} Prinn, R. G.,
   \& Barshay, S. S. 1977, Science, 198, 1031
\bibitem[Richardson et al.(2013)]{richardson13} Richardson, M. R.,
   Moses, J. I., Line, M. R., et al. 2013, 44th Lunar and Planetary Science Conference, 
   held 18-22 March 2013 in The Woodlands, TX, 2678 
\bibitem[Rogers \& Seager(2010a)]{rogers10frame} Rogers, L. A.,
   \& Seager, S. 2010a, \apj, 712, 974
\bibitem[Rogers \& Seager(2010b)]{rogers10gj1214b} Rogers, L. A.,
   \& Seager, S. 2010b, \apj, 716, 1208
\bibitem[Rogers et al.(2011)]{rogers11} Rogers, L. A.,
   Bodenheimer, P., Lissauer, J. J., \& Seager, S. 2011, \apj, 738, 59
\bibitem[Saffe et al.(2005)]{saffe05} Saffe, C.,
   G\'omez, M., \& Chavero, C. 2005, A\&A, 443, 609
\bibitem[Sanz-Forcada et al.(2010)]{sanzforcada10} Sanz-Forcada, J.,
   Ribas, I., Micela, G. et al. 2010, A\&A, 511, L8
\bibitem[Sanz-Forcada et al.(2011)]{sanzforcada11} Sanz-Forcada, J.,
   Micela, G., Ribas, I., et al. 2011, A\&A, 532, A6
\bibitem[Schaefer \& Fegley(2009)]{schaefer09} Schaefer, L.,
   \& Fegley, B., Jr. 2009, \apj, 703, L113
\bibitem[Schaefer et al.(2012)]{schaefer12} Schaefer, L.,
   Lodders, K., \& Fegley, B., Jr. 2012, \apj, 755, 41
\bibitem[Seager et al.(2005)]{seager05} Seager, S.,
    Richardson, L. J., Hansen, B. M. S., et al. 2005, \apj, 632, 1122
\bibitem[Segura et al.(2007)]{segura07} Segura, A.,
   Meadows, V. S., Kasting, J. F., Crisp, D., \& Cohen, M. 2007, A\&A, 472, 665
\bibitem[Selsis et al.(2002)]{selsis02} Selsis, F.,
   Despois, D., \& Parisot, J.-P. 2002, A\&A, 388, 985
\bibitem[Shabram et al.(2011)]{shabram11} Shabram, M., 
   Fortney, J. J., Greene, T. P., \& Freedman, R. S. 2011, \apj, 727, 65
\bibitem[Sharp \& Burrows(2007)]{sharp07} Sharp, C. M.,
   \& Burrows, A. 2007, ApJS, 168, 140
\bibitem[Showman et al.(2009)]{showman09} Showman, A. P.,
   Fortney, J. J., Lian, Y., et al. 2009, \apj, 699, 564
\bibitem[Shporer et al.(2009)]{shporer09} Shporer, A.,
   Mazeh, T., Pont, F., et al. 2009, \apj, 694, 1559
\bibitem[Southworth(2010)]{southworth10} Southworth, J.
   2010, MNRAS, 408, 1689
\bibitem[Spiegel et al.(2010)]{spiegel10} Spiegel, D. S., 
   Burrows, A., Ibgui, L., Hubeny, I., \& Milsom, J. A. 2010, \apj, 709, 149
\bibitem[Stevenson et al.(2010)]{stevenson10} Stevenson, K. B., 
   Harrington, J., Nymeyer, S., et al. 2010, Nature, 464, 1161
\bibitem[Stevenson et al.(2012)]{stevenson12} Stevenson, K. B., 
   Harrington, J., Lust, N. B., et al. 2012, \apj, 755, 9
\bibitem[Stone(1976)]{stone76} Stone, P. H. 
   1976, in Jupiter, ed.~T.~Gehrels (Tucson, AZ: Univ.~Arizona Press), 586
\bibitem[Tinetti et al.(2012)]{tinetti12} Tinetti, G.,
   Beaulieu, J. P., Henning, T. et al. 2012, Exp. Astron., 34, 311
\bibitem[Torres et al.(2008)]{torres08} Torres, G.,
   Winn, J. N., \& Holman, M. J. 2008, \apj, 677, 1324
\bibitem[Venot et al.(2012)]{venot12} Venot, O.,
   H\'ebrard, G., Ag\'undez, M., et al.~2012, A\&A, 546, A43
\bibitem[Visscher(2012)]{visscher12} Visscher, C., 2012, \apj, 757, 5
\bibitem[Visscher et al.(2006)]{viss06} Visscher, C.,
   Lodders, K., \& Fegley, B., Jr. 2006, \apj, 648, 1181
\bibitem[Visscher et al.(2010a)]{viss10rock} Visscher, C.,
   Lodders, K., \& Fegley, B., Jr. 2010a, \apj, 716, 1060
\bibitem[Visscher et al.(2010b)]{viss10co} Visscher, C.,
   Moses, J. I., \& Saslow, S. A. 2010b, Icarus, 209, 602
\bibitem[Visscher \& Moses(2011)]{visscher11} Visscher, C.,
   \& Moses, J. I. 2011, \apj, 738, 72
\bibitem[von Braun et al.(2012)]{vonbraun12} von Braun, K.,
   Boyajian, T. S., Kane, S. R.,  et al. 2012, \apj, 753, 171
\bibitem[Weiss et al.(2013)]{weiss13} Weiss, L. M.,
   Marcy, G. W., Rowe, J. F., et al. 2013, \apj, 768, 14
\bibitem[Woods \& Rottman(2002)]{woods02} Woods, T. N.,
   \& Rottman, G. J. 2002, in Atmospheres in the Solar System: Comparative 
   Aeronomy, ed. M. Mendillo, A. Nagy, \& J. H. Waite (Washington, DC: American 
   Geophysical Union), 221
\bibitem[Yung \& DeMore(1999)]{yung99} Yung, Y. L.,
   \& DeMore, W. B. 1999, Photochemistry of Planetary Atmospheres, (Oxford: Oxford Univ. Press)
\bibitem[Zahnle et al.(2009)]{zahnle09sulf} Zahnle, K., 
   Marley, M. S., Freedman, R. S., Lodders, K., \& Fortney, J. J. 2009, \apj, 701, L20





\end{thebibliography}
\end{document}